\documentclass[twocolumn]{aastex631}

\newcommand\starkiller{\texttt{starkiller}}
\newcommand{\kms}{km\,s$^{-1}$}

\graphicspath{{./}{figs/}}

\received{November 22, 2024}

\submitjournal{AJ}

\shorttitle{Starkiller: subtracting stars from IFU data}
\shortauthors{Ridden-Harper et al.}

\begin{document}

\title{\texttt{Starkiller}: subtracting stars and other sources from IFU spectroscopic data through forward modeling}

\correspondingauthor{Ryan Ridden-Harper}
\email{ryan.ridden@canterbury.ac.nz}

\author[0000-0003-1724-2885]{Ryan Ridden-Harper}
\affiliation{School of Physical and Chemical Sciences -- Te Kura Mat\={u}, University of Canterbury, Private Bag 4800, Christchurch 8140, Aotearoa New Zealand}

\author[0000-0003-3257-4490]{Michele T. Bannister}
\affiliation{School of Physical and Chemical Sciences -- Te Kura Mat\={u}, University of Canterbury, Private Bag 4800, Christchurch 8140, Aotearoa New Zealand}

\author[0000-0003-1955-628X]{Sophie E. Deam}
\affiliation{School of Physical and Chemical Sciences -- Te Kura Mat\={u}, University of Canterbury, Private Bag 4800, Christchurch 8140, Aotearoa New Zealand}
\affiliation{Space Science and Technology Centre, School of Earth and Planetary Sciences, Curtin University, Perth, Western Australia 6845, Australia}

\author[0000-0001-5344-8069]{Thomas Nordlander}
\affiliation{Research School of Astronomy and Astrophysics, Australian National University, Canberra, ACT 2611}
\affiliation{ARC Centre of Excellence for All Sky Astrophysics in 3 Dimensions (ASTRO 3D), Australia}

\begin{abstract}

We present \texttt{starkiller}, an open-source Python package for forward-modeling flux retrieval from integral field unit spectrograph (IFU) datacubes.
\starkiller\ simultaneously provides stellar spectral classification, relative velocity, and line-of-sight extinction for all sources in a catalog, alongside a source-subtracted datacube.
It performs synthetic difference imaging by simulating all catalog sources in the field of view, using the catalog for positions and fluxes to scale stellar models, independent of the datacube. 
This differencing method is particularly powerful for subtracting both point-sources and trailed or even streaked sources from extended astronomical objects. 
We demonstrate \starkiller's effectiveness in improving observations of extended sources in dense stellar fields for VLT/MUSE observations of comets, asteroids and nebulae. 
We also show that \starkiller\ can treat satellite-impacted VLT/MUSE observations.
The package could be applied to tasks as varied as dust extinction in clusters and stellar variability; the stellar modeling using \textit{Gaia} fluxes is provided as a standalone function.
The techniques can be expanded to imagers and to other IFUs.

\end{abstract}


\section{Introduction} \label{sec:intro}


Integral field unit spectrographs (IFUs) combine the strengths of imaging and spectroscopy: both spatial and spectral resolution for every spaxel in a given field of view.
Coupled with adaptive optics, this makes them a mainstay of extended-source observational astronomy --- serving a variety of communities including stellar evolution, star clusters, Galactic and extragalactic science.
Multiple generations of IFUs have been built for facilities around the world, from WiFeS on the ANU 2.3 m at Siding Spring \citep{Dopita:2010}, to MUSE on ESO's UT4 of the VLT \citep{Bacon:2010} and JWST's NIRSpec \citep{Boker:2022}.
Indeed, every proposed thirty-metre-class optical facility (the ELTs) includes an IFU as a first-light instrument. 
However, IFU data currently presents challenges for some science cases.

\subsection{Case 1: a background extended target with foreground stellar fluxes e.g. a nebula}

In the situation where an extended source such as a nebula or low-surface-brightness galaxy has a foreground and/or background stellar field, the data will be acquired with sidereal tracking, and all sources will have circular point-source point-spread functions (PSFs).
The density of the stellar field can limit the potential inference of the extended source's morphology and composition.
The PampelMUSE package \citep{Kamann:2013} provides precise PSF fitting for photometry in dense stellar fields in IFU data. 

\subsection{Case 2: a foreground extended target with background trailed stellar fluxes e.g. a comet}

In contrast to other spectroscopic modes, IFUs have seen comparatively little use by the Solar System small-bodies community, whose targets are frequently unresolved sources.
A common use of IFUs in Solar System studies has been atmospheric characterisation of bright planets (e.g. instruments such as Gemini/NIFS and VLT/SINFONI \citep{Simon:2022}).
However, IFUs are ideal for compositional studies of small Solar System bodies with activity creating extended comae --- but like all Solar System targets, these worlds move across the sky.
The requirement of non-sidereal telescope tracking to increase signal-to-noise streaks all background sources.
For imaging, median stacking can adequately remove the streaked stellar signal, but this often limits any characterization of time-varying phenomena. 
Studies of active minor planets are historically kept to areas of sky with low stellar density, with observation programs paused when targets traverse the Galactic plane (e.g. the 2022-23 DART mission post-impact followup campaign; \citet{Moskovitz:2023arXiv231101971M, Kareta:2023arXiv231012089K}).
For imaging in dense stellar fields, image subtraction is now a robust technique that aids both minor planet detection, and characterization such as lightcurve studies, but it is not yet used frequently for IFUs.

\subsection{Case 3: a celestial target with foreground streaks e.g. satellite streaks or trailed asteroids}

As the industrialization of near-Earth space increases, the astronomical communities also face the advent of streak-smeared data.
The accelerating rate of satellite constellation emplacement into low-Earth orbit (LEO) means $>6400$ have been launched, with $>5800 $ operationally in place as of April 2024\footnote{\url{https://planet4589.org/space/con/conlist.html}}; at least 20,000-100,000, with potential for around a million LEO satellites, will be in place in the mid-2030s \citep{Walker:2021,Walker:2022,Falle:2023}.
The population will then need to be maintained at that level by ongoing launches.
These satellites produce an industrially-caused environmental impact on astronomical observations: out-of-focus streaks of reflected Solar flux if they traverse the field of view, obscuring science targets, during an exposure.
While the probability of satellite streaks affecting the smaller fields of view of IFUs is lower than that for the massively wide-field imager of the Vera C. Rubin Observatory, the probability of effects on astronomical imaging is already non-zero \citep{Walker:2021, Michałowski:2021,Walker:2022}.
IFUs typically acquire longer integrations on target than e.g the 15s/30 s exposures of the LSST, so there is a different scope for adverse impact on IFUs. 
We demonstrate example MUSE data impacts in \S~\ref{sec:satellites}.
The environmental impacts will only continue to increase, particularly in the era of ELTs. 
By the expected European ELT first light, a steady-state LEO population kept by industry at some 30,000 satellites could exist; it is certainly on track to exceed 15,000.

For distant celestial targets acquired with sidereal tracking, near-field minor planets, particularly asteroids at geometries outside of quadrature, will also trail to varying degrees. 
The data on both types of astronomical object will be of interest to different communities.

\subsection{The \starkiller\ package}

\begin{figure*}
\centering
\includegraphics[width=\textwidth]{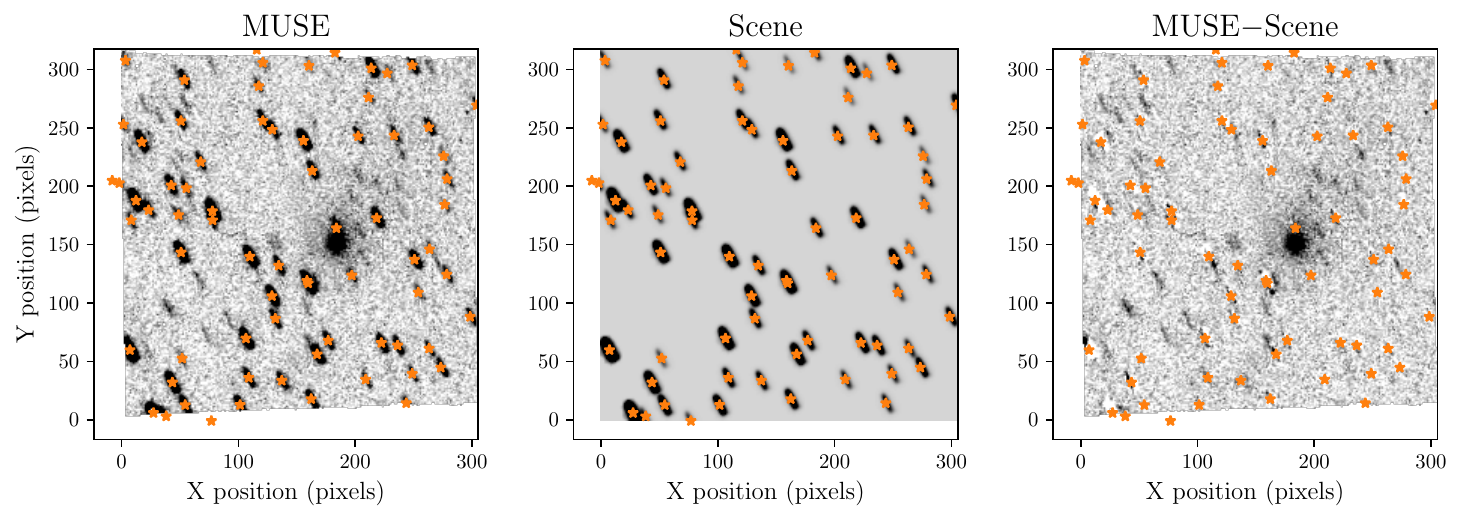}
\caption{
\textbf{Left:} Reduced MUSE datacube at $7000\,\rm \AA$ of 2I/Borisov at $b=26\degr$ on 2020-03-19, with position-corrected \textit{Gaia} DR3 stars shown as orange stars. 
\textbf{Center:} Scene constructed by \starkiller\ on \textit{Gaia}-listed stars in the field of view. 
\textbf{Right:} The scene-subtracted cube at $7000\,\rm \AA$. 
Subtraction quality varies between sources, influenced primarily by the closeness of the model spectra, and crowding. 
Sources that are not present in the \textit{Gaia} DR3 catalog are not subtracted, and are therefore present unaltered in the subtracted scene.
\label{fig:subtraction}}
\end{figure*}

Here we present a new open-source forward-modeling approach to removing stellar flux and satellite streaks from IFU datacubes, regardless of whether they are trailed or round.
We model the flux of the stars in the field that are identifiable in a source catalog, via stellar atmosphere models.
As the flux can be streaked or circular, a trailed point-spread function (PSF) is constructed to fit the stellar PSF of each datacube.
We apply location-appropriate dust extinction and find the best-match stellar spectra for each star. 
The streaks then form a simulated data cube, which is subtracted from the original (e.g. Fig.~\ref{fig:subtraction}). 
This approach allows us to replicate traditional difference imaging to remove background stars without the need for a reference data cube.
All magnitudes are assumed to be AB, and fluxes in terms of F$_\lambda$ ($\rm erg/s/cm^2/\AA$) unless otherwise stated. 

Our approach is inherently generalisable between instruments. 
We use VLT/MUSE here as a case study, with two extended small Solar System bodies, a nebula (\S~\ref{sec:sidereal}), and a satellite streak that pass over a blazar (\S~\ref{sec:satellites}) as examples. 
With suitable stellar models across the appropriate wavelength range, and a well-characterized instrumental PSF, \starkiller\ can be extended to other IFUs (\S~\ref{sec:other_ifus}).
Pull requests are welcome\footnote{\url{https://github.com/CheerfulUser/starkiller}}.

\section{Data}
 
\subsection{Example IFU: VLT/MUSE}
\label{sec:muse}

The ESO Very Large Telescope's Multi Unit Spectroscopic Explorer \cite[VLT/MUSE;][]{Bacon:2010} is a panoramic integral field unit spectrograph covering 4000\AA{}--9300\AA{}, on the 8.2 m UT4 optical telescope at Paranal, Chile. 
In wide-field mode, the field of view (FOV) is $1\arcmin \times 1\arcmin$, ideal for imaging extended sources. The optionally adaptive-optics corrected light is split equally and fed to 24 individual spectrographs (integral field units) \citep{Bacon:2010}.

We used contrasting MUSE datasets acquired with both sidereal and non-sidereal tracking for our primary development and testing of \starkiller. 
For sidereal data, we use two example selections from the ESO Archive: the planetary nebula NGC 6563, and satellite-impacted observations from 2021-22 of blazar WISEA J014132.24-542751.0 (J0141-5427).

We use two non-sidereal MUSE datasets, one from each MUSE mode, each of which present different data-analysis challenges.
The first is of the interstellar comet 2I/Borisov, observed on 16 epochs in 2019-2020 \citep{Bannister:2020, Deam:2024}. 
These data have increasingly dense stellar backgrounds, as 2I/Borisov moved from 49\degr\ outside the Galactic plane to within the plane after it passed perihelion.
The stars are streaked up to $\sim 20\arcsec$ at high galactic latitude, with the shortest $\sim 4\arcsec$ at low galactic latitude.
2I/Borisov has a compact coma, entirely contained within at most 46\arcsec\ diameter, and thus fully within the MUSE WFM FOV. 

Our other non-sidereal dataset is from followup of the NASA DART mission's impact on 2022 September 26 of the near-Earth asteroid moon Dimorphos \citep{Opitom:2023,Murphy:2023}.\footnote{Note that the moon is not resolved in these observations.} 
The resulting debris formed extended and time-varying structures over the following month relative to the bright parent body in the system, Didymos.
While \citet{Opitom:2023,Murphy:2023} acquired 11 epochs, only the last three have dense stellar backgrounds, when Didymos moved onto the Galactic plane.
These data were acquired at very rapid motion rates tracked on Didymos due to its geocentric proximity of only 0.08--0.09 au, which produced longer stellar streaks than those in the 2I/Borisov data; most streaks are not completely enclosed in the FOV.
The alignment of the FOV was offset on Didymos and rotated $90^{\circ}$ with respect to that of the 2I data, with the data acquired in MUSE's $8\arcsec\times8$\arcsec\ narrow-field mode with AO. 


\subsection{Catalog: Gaia}
\label{sec:gaia}
To identify stars within each cube, we use the \textit{Gaia} DR3 source catalog for star positions and brightness. 
\textit{Gaia} provides an all-sky catalog of sources with a limiting magnitude of 22 and a saturation magnitude of $\sim3$ \citep{GAIA,GAIADR3,Babusiaux2023}. 
The precise positions of \textit{Gaia} sources assists in sub-spaxel alignment of stars, while the broad \textit{G} band magnitudes are ideal for scaling model spectra flux. 

\begin{figure}
    \centering
    \includegraphics[width=\columnwidth]{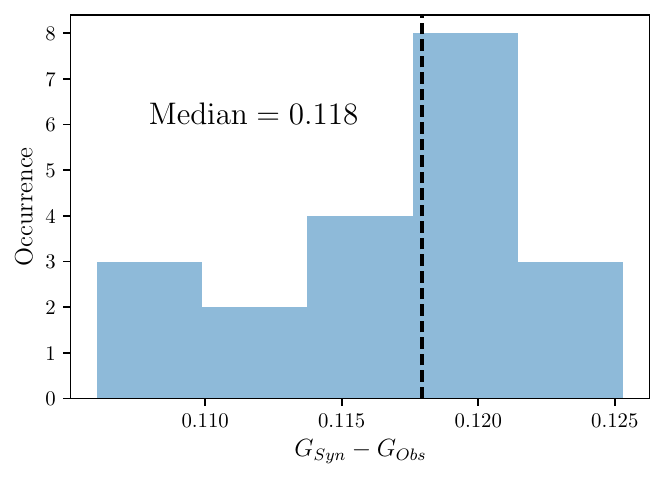}
    \caption{Distribution of differences between the Synthetic G band AB magnitudes ($G_{Syn}$) and the observed G band Vega magnitudes ($G_{Obs}$) for DA white dwarf calibrators. 
    The offset between the two magnitudes is primarily due to the differences in magnitude systems. 
    There is also a small calibration offset, which is described in \citet{Axelrod2023}.}
    \label{fig:gaia_cal}
\end{figure}

As \starkiller\ assumes all magnitudes are in the AB system, we must apply a correction to the \textit{Gaia} DR3 magnitudes, which are presented in the Vega system. 
We compute the correction by following the same procedure as \citet{Axelrod2023}, comparing observed \textit{Gaia} G magnitudes (Vega) to synthetic AB magnitudes in the G filter for 17 well-calibrated DA white dwarfs \citep{Narayan2019}. 
We take the median offset between the synthetic and observed magnitudes to be the correction factor to map \textit{Gaia} G (Vega) to \textit{Gaia} G (AB). 
Using the median in Fig.~\ref{fig:gaia_cal}, the correction becomes:  
\begin{equation}
G_{AB}=G_{Vega}+0.118
\end{equation}

\subsection{Stellar Atmosphere Models}
\label{sec:atmospheres}

High-quality stellar atmosphere models are essential for reliably representing stars within the data cubes. 
For accurate spectral matching, we need a diverse set of model spectra that span wide ranges in effective surface temperature T$_{\rm eff}$, surface gravities $\rm log(g)$, and metallicities. 
In this initial demonstration of \starkiller\ we also prioritize ease of use, so we restrict ourselves to smaller spectral libraries that can be installed alongside the base code. 
In order to best represent the spectral types of any star, we provide multiple models and implementation pathways between these models. 

Primarily, we use the \citet{Castelli2003} stellar atmosphere models (CK models) as a basis for stellar spectra comparisons. 
We choose the STScI subsection of the total CK atlas to cover the key range of spectral types\footnote{\href{https://www.stsci.edu/hst/instrumentation/reference-data-for-calibration-and-tools/astronomical-catalogs/castelli-and-kurucz-atlas}{https://www.stsci.edu/hst/instrumentation/reference-data-for-calibration-and-tools/astronomical-catalogs/castelli-and-kurucz-atlas}}.
We also include the ESO stellar spectra library\footnote{\href{https://www.eso.org/sci/facilities/paranal/decommissioned/isaac/tools/lib.html}{https://www.eso.org/sci/facilities/paranal\\ /decommissioned/isaac/tools/lib.html}} which use the \citet{Pickles1998} stellar spectra, supplemented with corrections from \citet{Ivanov2004}. 
(While there is a MUSE-specific library, it only contains 35 stellar spectra and is restricted to the MUSE wavelengths, so we do not use it here).
We include a partial grid of medium-resolution ($R = 20\,000$) sampled fluxes from the MARCS grid \citep{Gustafsson2008}. 
We have selected spectra covering the most common types of late-type stars, with $T_{\rm eff} = 3000$--8000\,K and $\log\,g$ between $-0.5$ and $+5.0$, in a pattern that broadly follows the main sequence and red giant branch. 
The models have $\rm [Fe/H] = -0.5$, 0.0 and $+0.5$, using solar-scaled abundances except that $[\alpha/{\rm Fe}] = +0.2$ when $\rm [Fe/H] < 0$. 
For simplicity, a single value of the micro-turbulence value $v_{\rm mic} = 2$\,\kms\ was selected. 
Models of both plane-parallel and spherical geometry (assuming stellar masses of one solar mass) are included where available. 
Finally, we include the PoWR grid of OB-type synthetic spectra \citep{Hainich2019}, called \textit{OB-i}. 
Specifically, we include the solar-metallicity grid that covers much of the parameter space $T_{\rm eff} = 15$--56\,kK and $\log\,g = 2.0$--4.4, broadly corresponding to the evolution of stars of roughly 7--60\,$M_\odot$.

\section{Analysis structure} 
\label{sec:style}

The process of analysis of \starkiller\ is generalized to operate on any optical IFU data that has the same header and HDU format as VLT's MUSE, including WCS information. 
We have minimized the amount of additional inputs required, with a strong preference towards self-determination of key information from the input datacube. 
The following sections outline the key steps we use in determining the stars within the field of view (FOV), and modeling those sources.

\subsection{Source catalog}
\label{sec:catalogue}

By default, we use the \textit{Gaia} DR3 catalog \citep{GAIADR3} as the source catalog. 
\starkiller\ obtains the \textit{Gaia} sources within a radius defined by the size of the IFU centered on the R.A. and Decl. provided in the input cube's header. 
We access the \textit{Gaia} DR3 catalog I/355/gaiadr3 through Vizier via \textsc{astroquery}. 
 
Alternatively, \starkiller\ also accepts user input source catalogs. 
The input catalogs must contain columns specifying the source ID (id), R.A. (ra), and Dec. (dec) in degrees, magnitude ($x$\_mag), and a filter designation for the SVO filter service\footnote{\href{http://svo2.cab.inta-csic.es/theory/fps/}{http://svo2.cab.inta-csic.es/theory/fps/} In the process of creating this designation, the Euclid filters were corrected within SVO as of 2023 June 20.} ($x$\_filt) where $x$ is the desired filter shorthand.

If multiple filters are specified for a single source, then the model spectra will be reshaped to match the magnitudes in all filters; unless the \texttt{key\_filter} argument is defined, identifying the filter to which to normalise the flux.

\subsection{WCS correction}
\label{sec:wcs}

As the purpose of \starkiller\ is to simulate spectral data cubes to provide simulated differenced cubes, precise positions of sources are essential: while this may be straightforward for siderally tracked observations, for those tracked non-sidereally, both the spatial WCS solutions and source identification require additional refinement. 
For MUSE, the WCS solution produced by the MUSE data reduction pipeline \citep{MPDAFsoftware} is propagated from the target coordinates defined in the VLT's observing block. 
For sidereally tracked observations, the spatial WCS solutions are effective: only minor corrections on the order of a spaxel are generally required.
In non-sidereal cases, the offset needed from the \citet{MPDAFsoftware} WCS is determined by the on-sky motion rate of the target, relative to the ephemeris timestamp choice when the VLT begins observation setup, modulo the typical MUSE acquisition time of 10-15 minutes.
For example, we find that the WCS solution can be offset by an arcminute in our 2I/Borisov data.
However, elongated sources then present a challenge for conventional source identifiers. 

To identify sources in the IFU, regardless of shape, we use clustering algorithms on an `image' constructed from a median stack of the input cube in wavelength space. 
We create a Boolean image with a percentile cut, selecting for spaxels that are brighter than the 90th percentile. 
We then label sources in the Boolean image with \texttt{scipy.ndimage.label}. 
Sources are then down-selected to retain the labeled objects with a total spaxel count between 0.01\% and 10\% of the total IFU spaxels\footnote{For MUSE, this limit corresponds to approximately 100 to 10000 spaxels.}. 
This cut limits contamination from sources that partially fall within the FOV (lower limit) and the background spaxels (upper limit). 
The center points of the labeled sources are taken to be the average $x$ and $y$ spaxel positions of spaxels for each source. 
We also extract initial guesses for the stellar PSF, by estimating the trailing length, trail angle, and the x and y spaxel extents. 

With sources identified in the image, we fit basic offsets from the star catalog to the image. 
Within \starkiller\ there are two methods to conduct an initial match, and a final method to refine matching sources with the catalog. 
The first and less reliable preliminary match method is to fit for for $x$, $y$, and rotational offsets, by minimizing the distance of the spaxel coordinates of the brightest stars in the catalog to the image sources. 
However, this method struggles for crowded fields, where the regions labeled as sources are composites of multiple sources and not representative of the PSF. 

The second preliminary match method, which is robust to crowding, assumes that the position angle on the sky has low error, and matches the catalog through shifts. 
In this method we first iterate through $x$ and $y$ offsets to the raw catalog spaxel coordinates from -100 to 100 spaxels, in steps of 10 spaxels. 
For each shift, we generate a simulated image by convolving sources within the image bounds with the profile of the median labeled source, and subtract this from the labeled image, which is altered such that sources are represented by 1 and all other spaxels are NaN. 
The pair of $x$ and $y$ offsets that provide the smallest residual are then taken as the starting parameters, to minimize the residual between the two images with \texttt{scipy.minimize}. 
This method is the default method used in \starkiller\ and provides close matches, even in crowded fields. 
For point sources this match is within 1~spaxel; the uncertainty does increase in crowded fields with high source elongation.

Following the initial catalog match, the WCS correction is then refined through PSF fitting. The creation of the PSF that is used in this routine is outlined in \S~\ref{sec:psf}. 
We take the $x$ and $y$ positions of the calibration sources found through PSF fitting, and compare those with the shifted catalog positions. 
Following a method similar to the first catalog matching method discussed, we minimize the distance between the shifted catalog positions and the PSF positions through $x$ and $y$ shifts as well as a rotation $\theta$ around the image center. 
This refined PSF shift matches observed sources with the source catalog positions to a sub-spaxel precision. For trailed sources, we consider the center points as defined by the PSF to be the observed position.
 
With these methods developed for \starkiller, we are able to correct for any errors present in the spatial WCS solution due to challenging observational conditions, such as non-sidereal tracking. 
While effective, these methods may lose reliability in highly crowded fields or with very elongated sources.
We discuss this further in \S~\ref{sec:limitations}.

\begin{figure}
    \centering
    \includegraphics[width=\columnwidth]{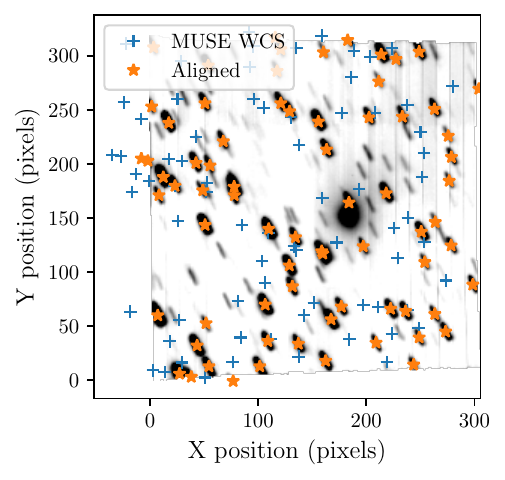}
    \caption{Comparison of the \textit{Gaia} DR3 star position using the MUSE WCS (blue +), and the \starkiller\ corrected positions (orange star). Through the \starkiller\ reduction process sources are realigned assuming a linear offset, plus rotation. In all instances sources are aligned to be at the center of their corresponding streaks. The alignment method used by \starkiller\ has proven to be robust to source elongation and crowding.}
    \label{fig:alignment}
\end{figure}

\subsection{Isolating sources}
Identifying isolated sources in non-sidereal data presents an interesting challenge as the sources may be streaked to any length, and aligned on any angle. 
As with the WCS correction, we adopt a 2 stage approach to identifying isolated sources, where we use the initial PSF approximations in stage 1, which we then refine with the preliminary PSF in stage 2. 

In stage 1, we rotate the coordinates of the catalog sources and image according to the estimated trail angle such that the trails are vertical. 
A source is then considered to be isolated if it is more than 8 spaxels from a neighbor in the $x$ direction (PSF minor axis), and separated by more than 1.2 times the total trail length in the $y$ direction (PSF major axis). 
We also incorporate magnitude information into the isolation criterion. 
If nearby sources are at least 2 magnitudes fainter, it is assumed that their contribution is small and therefore ignored when calculating source distances. 
The isolated sources identified in this process are used to generate the first iteration of the PSF.

Stage 2 relies on a PSF being defined to determine a refined calibration source list. 
To identify isolated sources using the PSF, we check overlaps between Boolean masks, created by placing the PSF at each source position where spaxels must contribute more than $1\times 10^{-5}$ to the total PSF. 
If any two masks contain overlapping points, the sources are considered to be overlapping.
By using the PSF information, we can reliably identify isolated sources, regardless of trail length or orientation. 
Sources we identify as ``isolated" through this process are then used as the final calibration sources, from which the final PSF and WCS correction are calculated.

\subsection{PSF modeling}
\label{sec:psf}

A precise PSF is key to creating accurate simulations of data cubes. 
We developed \starkiller\ to model PSFs for static sources and elongated sources, generated from sidereal and non-sidereal tracking. 
Our PSF module is built from the PSF module within \texttt{TRIPPy} \citep{Fraser:2016}, which downsamples a high resolution PSF to the image resolution. 
Alongside the Moffat profile PSF used in \texttt{TRIPPy}, we also incorporate a Gaussian profile PSF, and a data-generated PSF; we refer to the latter as the ``data PSF". 

As constructed in \texttt{TRIPPy}, a trailed PSF can be reliably modeled by a non-trailed point source PSF profile convolved with a line model. 
Fitting trailed PSFs therefore requires fitting the PSF profile parameters alongside the line model which incorporates trail length and angle, which we choose to be the angle measured counter-clockwise from the $x$ axis. 
In \starkiller\ the non-trailed point source PSF profile can be either Moffat, or symmetric 2D Gaussian, and is defined by fitting to calibration sources that are considered ``isolated'' and brighter than a user-defined calibration magnitude limit. 

When constructing the model PSF, we simultaneously fit all parameters for the profile and line element, including small positional offsets. 
The models are generated at 10 times the spatial resolution and downsampled to the data resolution. 
While the PSF is dependent on wavelength, we find that these variations are small for the highly streaked stars in MUSE data. 
Therefore, to optimize the signal-to-noise of the calibration sources, \starkiller\ fits a single PSF using the median stack of all wavelengths. 

A further complication to the PSF fitting process is variability in the atmospheric seeing. 
For sidereal tracking, variations in seeing contribute evenly to the total PSF throughout the exposure. 
However, for non-sidereal tracking, the trailed PSF becomes a record of the seeing variability during the exposure, where each segment of the trail may be constructed by different seeing conditions to another section. 
A trailed PSF may therefore be quite different from a distribution that can be readily modeled by a simple elongated Moffat, or Gaussian profile. 

\begin{figure}
    \centering
    \includegraphics[width=\columnwidth]{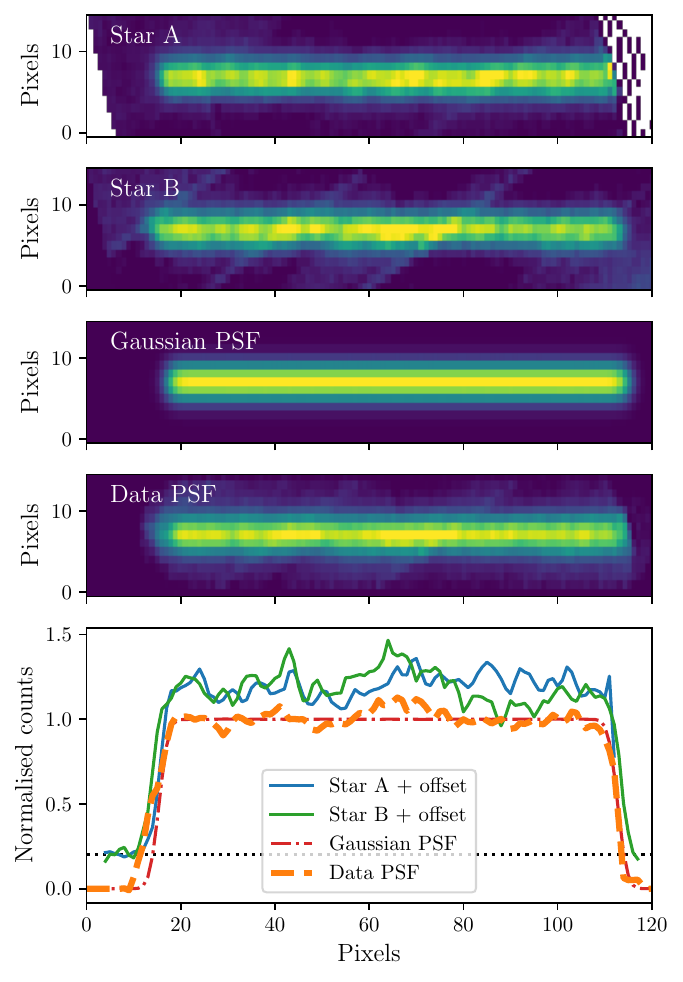}
    \caption{Seeing variability of highly elongated stars in MUSE observations of 2I/Borisov on 2019-Dec-31. 
    Stars A \& B, as shown in the top two panels, are the calibration stars used by \starkiller\ in this reduction, followed by their best-fitting Gaussian and data PSFs. 
    The stars are highly elongated, with a fitted streak length of 98~spaxels. 
    Over the course of this 600~s exposure, the star trails exhibit a non-uniform brightness profile along their major axis, which causes the stars to be poorly represented by a simple elongated Moffat or Gaussian profile. 
    The bottom panel shows the normalized cross section along the major axis for: stars A \& B (blue and green solid lines respectively); the best fitting Gaussian PSF (red dash-dot line); and the Data PSF (orange dashed line). 
    The Data PSF is able to capture some of the subtle structure imposed on the PSF from seeing variability, which leads to better subtractions.}
    \label{fig:scintillation}
\end{figure}

An example of highly trailed stars can be seen in Fig.~\ref{fig:scintillation}. 
Stars A \& B\footnote{Star~A:~Gaia~3462487872509843200. B:~Gaia~3462487666351399296.} (Fig.~\ref{fig:scintillation}, top two panels) were observed alongside 2I/Borisov with MUSE on 2019-12-31 during a 300~s exposure. 
The high slew rate throughout these observations created elongated stellar PSFs that were $\sim 100$~spaxels long. 
In this extreme case, variability can be clearly seen along the lengths of the stars --- it cannot be replicated in a simple elongated Gaussian (or Moffat).
While more complex models could be constructed that vary across the length of the trailed PSF, in \starkiller\ we instead create a data PSF: we normalize and average together the trails of all calibration stars in the data cube. 
While this simple approach can introduce noise structures into the PSF, such as seen in Fig.~\ref{fig:scintillation} (Data PSF), it reliably captures the seeing variability. 
In the major axis cross sections shown in the lower panel of Fig.~\ref{fig:scintillation}, the Gaussian PSF profile fails to capture the variability that is largely shared between the stars, which is present in the data PSF. 

In constructing the data PSF, \starkiller\ must only use stars that are well contained within the data cube. 
Therefore, alongside the magnitude limit and isolation requirement, we introduce a ``containment" requirement. 
We test PSF containment by creating individual images for all catalog stars by convolving an image containing their spaxel position with the model PSF function.
The implanted PSFs are then summed over the extent of the datacube to give a containment fraction. 
To ensure the data PSF is representative of the entire PSF, we only include sources with containment fractions $>95\%$ in its construction. 
In cases where there are few suitable calibration stars, the data PSF may be biased to those stars, and therefore not a fair representation of all sources. 

\begin{figure*}
\centering
\includegraphics[width=\textwidth]{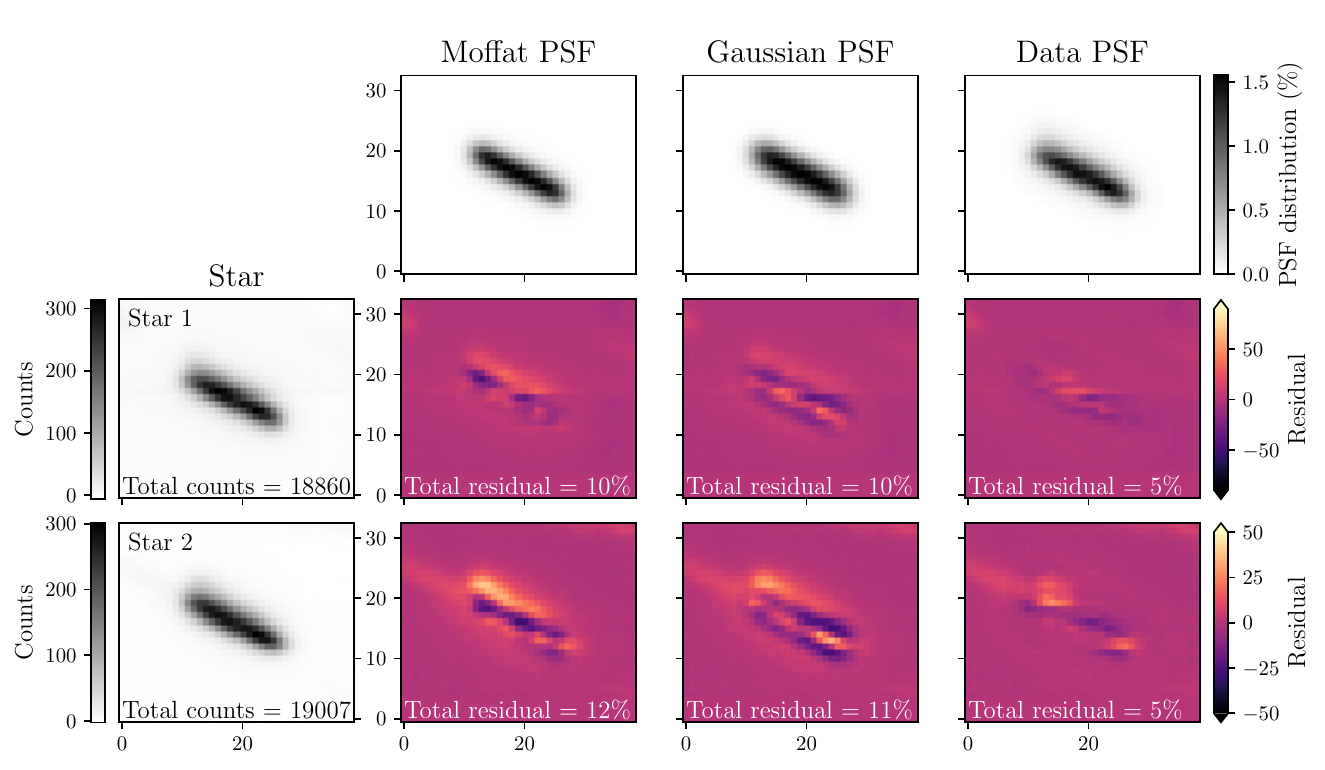}
\caption{Example of the 3 PSF models --- Moffat, Gaussian, and Data (top row) --- alongside two field stars in a 2I/Borisov datacube observed on 2020-03-19 (left column), and their respective residuals (center). 
While each PSF appears similar, the subtle differences become clear in the fit residuals. 
Neither the Moffat or Gaussian profiles provide a reliable representations of the stars in this datacube, or in most other streaked MUSE datacubes. 
However, the data PSF is a much closer match, lacking the poor subtraction patterns of the Moffat and Gaussian.
We attribute the complex PSF that significantly varies along the streak to atmospheric scintillation throughout the 600 s exposure. 
\starkiller\ will begin modeling the PSF as either Gaussian or Moffat; however, if the initial model is significantly different from the data PSF, the data PSF is then used exclusively in spectra extraction and scene generation. 
\label{fig:psf_fit}}
\end{figure*}

While spatial variability in the trailed PSF is clearest in the highly elongated sources, it is still present in sources with shorter trails. 
In general, we find that the data PSF provides a more accurate representation of trailed sources in MUSE datacubes. 
In Fig.~\ref{fig:psf_fit}, we use all three PSF methods to model 2 stars from the MUSE 2I/Borisov datacube observed on 2020-03-1\footnote{Star 1: Gaia~5856950561179336704. Star 2: Gaia~5856950561179334912.}. 
While all methods greatly reduce the total counts, both the absolute residual and visual artifacts are lowest for the trailed data PSF fits at $\sim5\%$, while the other two methods have $\sim10\%$ residuals. 
For sidereally tracked data we find the data PSF outperforms the other models by $\sim15\%$ as seen in Fig.~\ref{fig:psf_fit_ps} in Appendix~\ref{sec:ps_comparison}.

Therefore, after constructing the data PSF, \starkiller\ will check for differences between the model PSF (either Moffat or Gaussian) and the data PSF: if the difference is large, then \starkiller\ will default to using the data PSF for spectral extraction and scene creation. 
If desired, this behavior can be disabled by setting the ``\texttt{psf\_preference}" option of \starkiller\ to ``\textit{model}".

The computed PSF is used by \starkiller\ to extract observed spectra through PSF photometry, and ultimately to model sources in the scene.

\subsection{Spectral matching}
\label{sec:spec_match}
We identify the best-match spectra for a star through correlation comparisons with the model stellar spectra. To avoid model confusion from noise spikes, or external emission lines, we apply an iterative sigma clipping and smoothing procedure to observed spectra before calculating model correlations. In this procedure we cut points that have gradient absolute values greater than 10$\sigma$ above the median. By default the spetcra are clipped for 3 iterations before undergoing smoothing with a Savitzky–Golay filter, as implemented in \text{scipy}. These treated spectra are then correlated with model spectra.

Correlation allows us to make a morphological comparison of the similarities between two spectra, without considering flux scaling. 
For each IFU spectra, we calculate the Pearson correlation coefficient using \texttt{scipy.stats.pearsonr} for all available model spectra, downsampled to the input IFU's spectral resolution.
This approach (rather than e.g. $\chi^2$) emphasises relative shapes and avoids concerns of normalisation.
Since we want the closest match, we take the corresponding model to the IFU spectra to be that which has the largest positive correlation $p$-value. 

In \starkiller\ we provide multiple pathways for spectral matching, using the range of model catalogs described in \S~\ref{sec:atmospheres}.
The model catalog used is determined by the \texttt{spec\_catalog} argument. 
By default \starkiller\ uses `ck', which checks against the CK models. If `ck+' is specified then \starkiller\ uses the temperature of the selected CK model to identify relevant high resolution spectra to compare against.
If the CK model temperature is $<8000$~K, we compare to a set of MARCS models within a $\pm500$~K temperature range of the input value. 
Similarly, if the CK temperature is $>15000$~K, we check against all models in the OB-i model list (see Sect.~\ref{sec:atmospheres}). 
This approach provides the most comprehensive stellar spectral matching in \starkiller.
Alternately, other catalogs can be selected: setting \texttt{spec\_catalog} to `ck' will restrict the spectral fitting to only the CK models, while `eso' will use the library of stellar spectra listed by ESO. 
If the base spectral models included in \starkiller\ are insufficient for the desired case, the spectral catalogs that are used can be readily altered. 

As extinction from interstellar dust can significantly reshape spectra, we must incorporate extinction in the template matching. For every template spectrum we create an extinction grid by applying the \citet{Fitzpatrick1999} extinction model with $R_V=3.1$ over the range $0\leq E(B-V)\leq4$ in steps of 0.01 using the \texttt{extinction} \citep{Barbary2016} and \texttt{PySynphot} \citep{pysynphot} packages. We then calculate the correlation of the extracted spectra with the grid of reddened models. The model spectrum with the highest correlation is then used to represent the source, and re-scaled to  match the catalog magnitudes. Through simulated recovery tests we find that this method is robust to noise, and has minimal degeneracy between spectral type and extinction.

\begin{figure*}
\centering
\includegraphics[width=\textwidth]{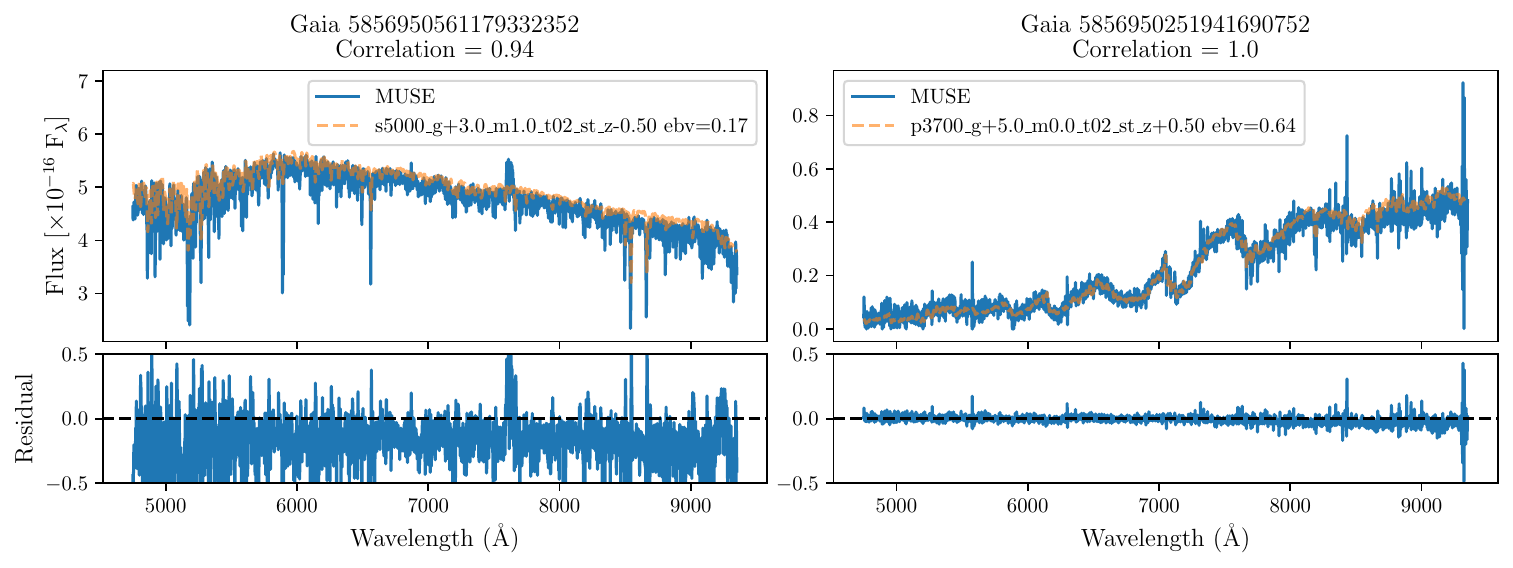}
\caption{Example spectral model match to two stellar spectra extracted from the trailed MUSE data cube in Fig.~\ref{fig:subtraction}. 
The MUSE spectra is shown by the solid blue line, and the best-fitting model's stellar spectra, which has been independently flux scaled by \textit{Gaia} G-band photometry, is shown as an orange dashed line. 
The best-matching model is taken to be the model with the highest Pearsons-r correlation, from a grid of spectra subject to a range of extinction values. An additional flux correction discussed in \S~\ref{sec:flux_correction} is also applied. 
The residuals of the model fits are shown in the lower panels. 
\label{fig:specfit}}
\end{figure*}

This spectral matching process provides the best approximation of the spectra from every star in the IFU. 
An example of one such fit is shown in Fig.~\ref{fig:specfit}. 
Our approach minimizes the input information from the IFU, thus reducing the likelihood that the stellar spectra are biased by the flux of other sources within the IFU, such as a foreground extended coma of a target comet.

\subsubsection{Optional output: Velocity matching}

An additional step to matching the fine detail of observed and model spectra is to apply corrections for any relative motion. 
While it is not used in the primary reduction procedure, \starkiller\ has a routine to identify the most likely Doppler shift to the observed spectrum. 
We calculate the likely shift by fitting Gaussian models to prominent absorption lines in stellar spectra: H$\beta$, H$\alpha$, Na D, and the Ca II triplet. 

\begin{figure*}
\centering
\includegraphics[width=\textwidth]{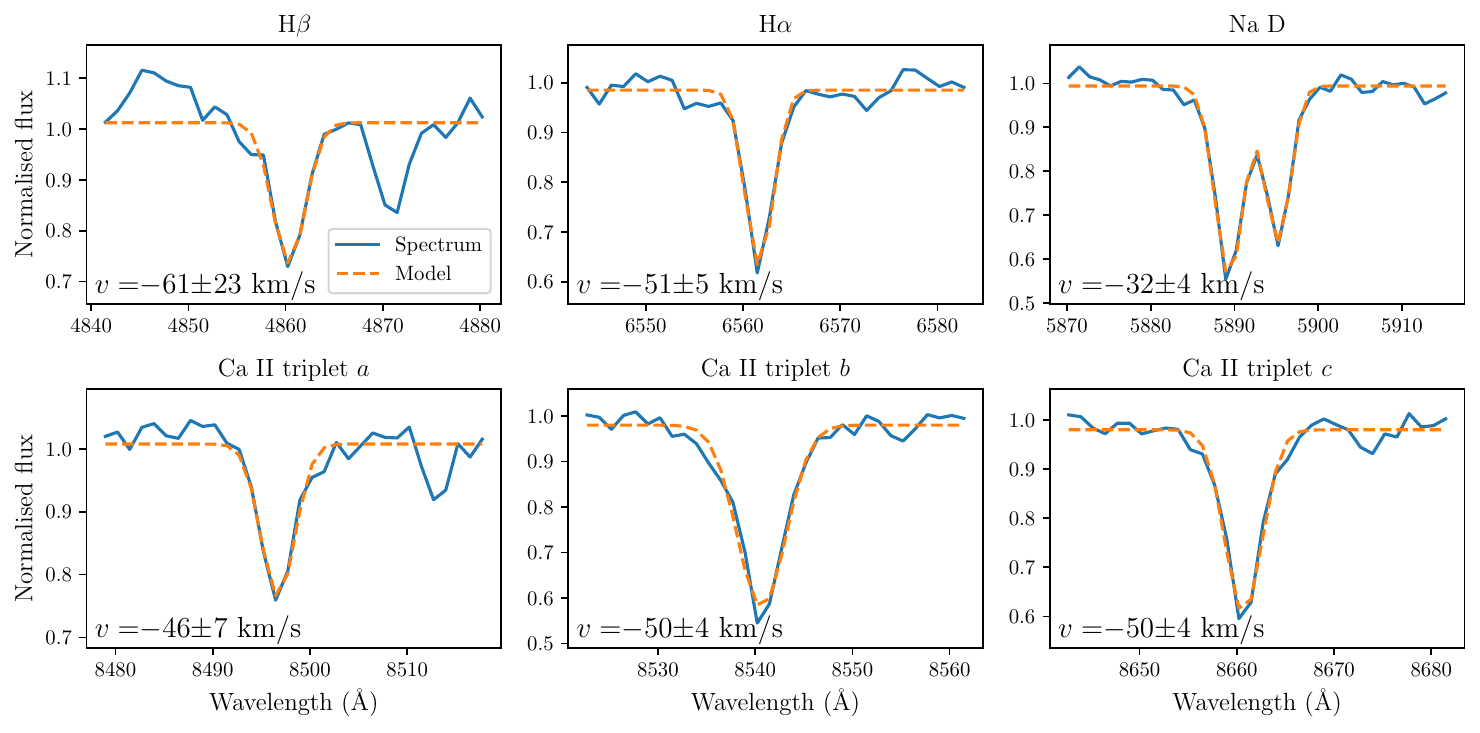}
\caption{
Doppler fits for Gaia~5856950561179332352 in the MUSE datacube of 2I/Borisov on 2020-03-19. 
\starkiller\ provides a rough relative velocity for stars with spectral types A--K by fitting Gaussian models to prominent absorption lines in stellar spectra: H$\beta$, H$\alpha$, Na D, and the Ca II triplet. 
Each line in the Ca II triplet is modeled independently and labeled as $a$, $b$, and $c$ in increasing wavelength. 
In this example, the error-weighted average velocity is $-46\pm2$~km/s.
\label{fig:doppler}
}
\end{figure*}

We fit the absorption lines with \texttt{astropy.modeling} independently with a single Gaussian plus an constant offset, with the exception of Na D which we fit with a double Gaussian model with a constant offset. 
The models are fit to a region of the spectrum $\pm20\;\rm\AA$ of the rest frame wavelength for the line. 
For each line, we normalize the spectra by the median flux value of a range $+20$ to $+40\;\rm \AA$ from the absorption line. 
The models are fit through \texttt{astropy}'s Levenberg-Marquardt algorithm and least squares statistic (LevMarLSQFitter) method, where the uncertainties are taken to be the square root of the diagonal elements covariance matrix. 
While we do not place bounds on the fit, we require the amplitude of the Gaussian models to be negative for it to be considered in the weighted average. 
The final velocity is taken to be the error-weighted average of the fit lines. 
An example of this method is shown in Fig.~\ref{fig:doppler} for star Gaia~5856950561179332352 from the MUSE datacube of 2I/Borisov on 2020-03-19. 
This method will only be effective for stars where the selected spectral features are prominent, such as spectral types A to K. 
However, it is not currently fully implemented in \starkiller, as the primary science cases have not to date been concerned with narrow lines.

\subsection{Flux correction} 
\label{sec:flux_correction}

The final adjustment that we make to the model spectra is to correct for any global trends in differences between the observed and model spectra. 
These wavelength-dependent differences may arise from issues with the MUSE flux calibration, or from consistently poorly matched model spectra. 
For \starkiller\ the origin of these global differences is inconsequential, as the primary goal is to replicate the IFU stellar spectra; for this, correcting for bulk trends as a function of wavelength is sufficient. 

We generate a wavelength-dependent flux correction by averaging the flux ratios of all sources with correlation coefficients $>0.9$. 
We further restrict the calibration sources to be brighter than the user-defined magnitude limit, to limit the influence of bad matches and noise on the flux correction. 
We then create a smoothed spline using a Savitzky-Golay filter using the \texttt{scipy} implementation, with a window size of $625\, \rm \AA$ (501~pixels) and a polynomial order of 3. 
The window size and polynomial order were chosen to avoid being biased by narrow features such as absorption lines, while being able to capture broader features alongside the continuum offset. 
We perform a 3 sigma clip on the difference between the flux ratios and smoothed spline, and refit the spline to further limit the influence of narrow line features. 

\begin{figure}
    \centering
    \includegraphics[width=\columnwidth]{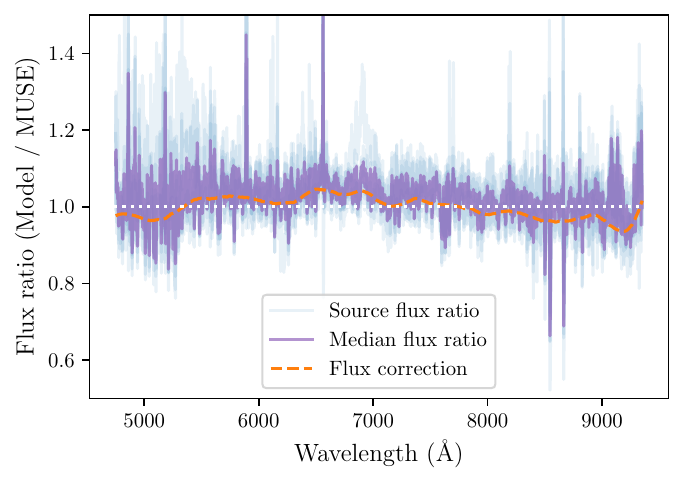}
    \caption{Ratio of the best fitting stellar models through the ck+ method to the MUSE spectra for each of the calibration sources from the IFU. 
    As each model spectrum is scaled by the catalog, in this case \textit{Gaia} DR3 magnitudes, any differences between the models can be a result of poor template matching, or variations in flux calibration. 
    Common trends that are present between the models (purple line) for the calibration sources are used to create a flux correction (orange dotted line). 
    Every model spectrum is divided by the flux correction, to bring them into greater alignment with the observed spectra. 
    As all variations occur close to 1, we can see that the flux calibration between \textit{Gaia} and MUSE is consistent.}
    \label{fig:flux_cor}
\end{figure}

As seen in Fig.~\ref{fig:flux_cor}, a characteristic correction curve emerges when examining the median flux ratio of all calibration sources. 
Our flux correction method is able to correct for the overall flux offset, and for larger features, such as the broad wiggles occurring along the spectrum, and the rapid rise after $9000\rm \, \AA$, while limiting bias from the poorly fit narrow lines. 
Dividing all model spectra by this flux correction brings them into closer alignment with the MUSE spectra. 
The model spectra shown in Figs.~\ref{fig:specfit} \& \ref{fig:nebula_specfit} have this flux correction applied.

While the method we present here is sufficient for \starkiller\ to reliably match model spectra to the observed MUSE spectra, it may not be reliable at distinguishing calibration errors from poor model fits.
As the origin of this offset is unknown, the flux correction is only applied to the model spectra to make them a closer match to the IFU data, therefore the calibration of the input data is not altered. 
It is worth noting that in general the flux calibration of MUSE spectra and \textit{Gaia} AB magnitude photometry are in agreement, with small deviations of $\sim10$\% occurring around a ratio of 1.

\subsection{Sidereal IFU data}
\label{sec:sidereal}

While \starkiller\ was developed for modeling non-sidereally-tracked IFU cubes, it is entirely capable of processing sidereally-tracked cubes. 
In these instances, the best-fitting trail length for the PSF profiles is $\sim1$, and the rotation angle becomes irrelevant. 

One complication can arise from datacubes with large extended structures, such as bright nebulae and galaxies, where the PSF profiles fail to fit correctly due to the underlying structure.
To limit the influence of this structure, the ``fuzzy" option can be set to \texttt{true} in \starkiller. 
In the ``fuzzy" mode, a ``fuzzymask'' will be created by applying \texttt{scipy.ndimage.label} to a Boolean image created by conditioning spaxel brightness in the datacube image based on the median spaxel brightness. 
If there are labels that occupy more than 40\% of the spaxels, it is considered for masking. 
The background is taken to be the label with the lowest median counts; the other labels are included in the fuzzymask. 
Sources within the fuzzymask will not be used for PSF creation or flux correction, even if they meet all other requirements to be considered a calibration source. 

\begin{figure*}
\centering
\includegraphics[width=\textwidth]{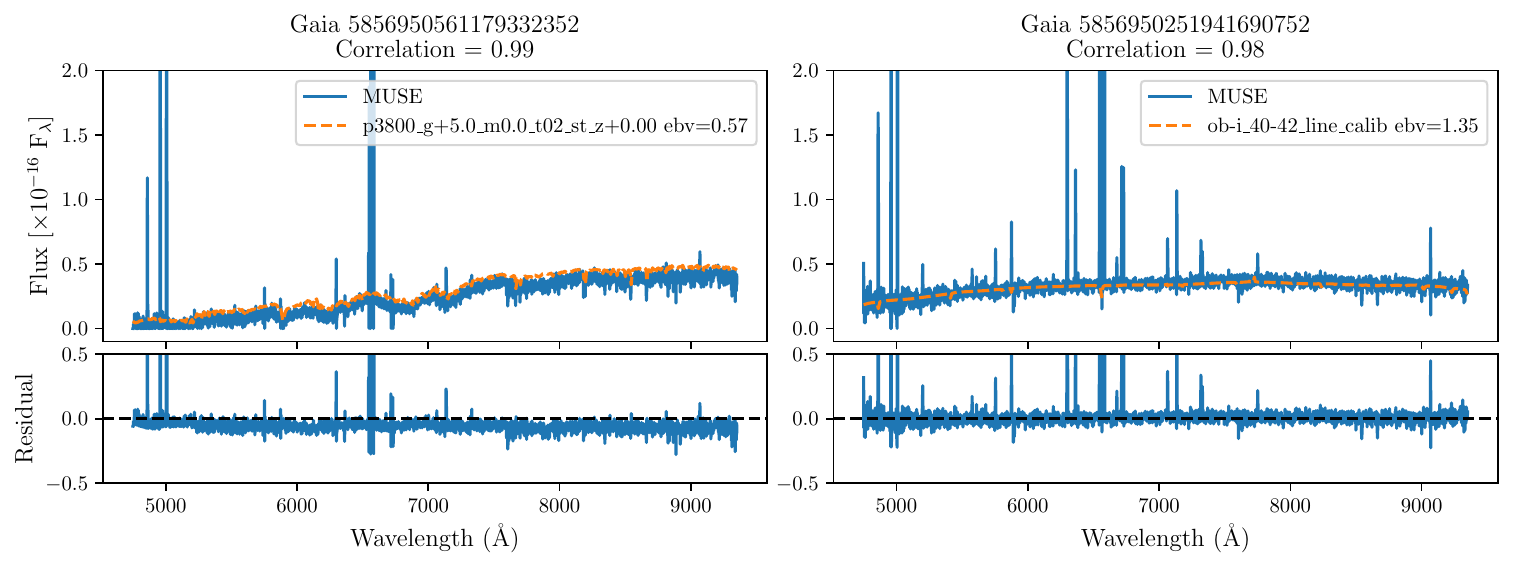}
\caption{Example spectral model match to two stellar spectra extracted from the MUSE data cube in Fig.~\ref{fig:nebula_subtraction}. 
The MUSE spectra is shown by the solid blue line, and the best-fitting model's stellar spectra which has been independently flux scaled by \textit{Gaia} G-band photometry, is shown as an orange dashed line (the same as in Fig.~\ref{fig:specfit}). 
Both of these sources are from within the nebula. The spectral extraction and modeling is robust to narrow features from external emission, or instrument noise, resulting in high model correlations.
\label{fig:nebula_specfit}}
\end{figure*}

\begin{figure*}
\centering
\includegraphics[width=\textwidth]{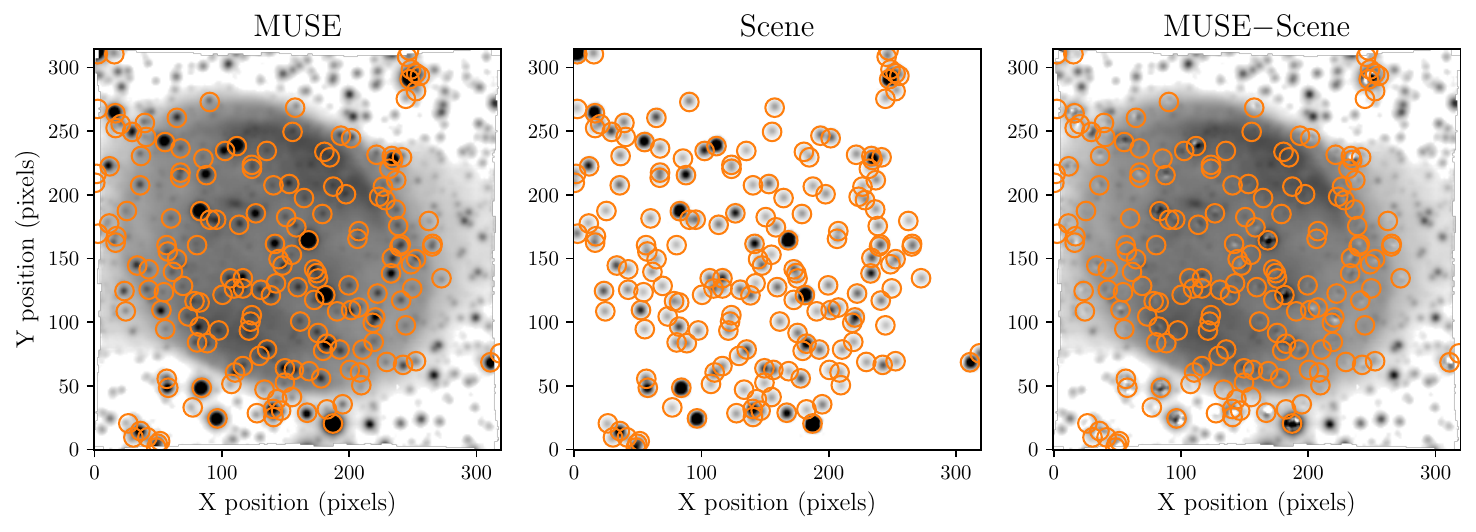}
\caption{\starkiller\ applied to the datacube of planetary nebula NGC~6563 (no AO) with position-corrected \textit{Gaia} DR3 stars shown as orange circles. 
(left) Mean stack of all wavelength for the reduced MUSE datacube. 
(centre) Scene constructed by \starkiller\ on \textit{Gaia}-listed stars in the field of view. 
(right) The median stack of all wavelengths for the scene-subtracted cube. 
The subtracted Gaia DR3 stars have residuals $<10\%$.  
\label{fig:nebula_subtraction}}
\end{figure*}

As an example, we apply \starkiller\ with the ``fuzzy" option enabled to a MUSE datacube of the planetary nebula NGC~6563, observed on 2018-08-22 without AO. 
An additional complication that we correct for in \starkiller, can arise from the presence of narrow features in the extracted spectra. While the spikes seen in Fig~\ref{fig:nebula_specfit} could result in low correlations with model spectra, the sigma clipping and smoothing algorithm that \starkiller\ applies results in high correlations between the extracted spectra and models. 
The full subtraction of NGC~6563 with a data-PSF (Fig.~\ref{fig:nebula_subtraction}) is successful in removing the majority of flux from sources that were identified by \textit{Gaia} with on average residuals of $<10\%$.

\subsection{Satellite detection and removal}
\label{sec:satellites}

By adapting the process that \starkiller\ uses to subtract stars, we are also able to identify spectra of satellites crossing an IFU, and attempt to remove them. 
Unlike stars, the magnitudes, SEDs, and on-sky locations of satellites are poorly constrained, making a full forward-modeling approach unachievable --- therefore, \starkiller\ relies entirely on the IFU data. 
In this prototype case, we focus on satellite detection and removal by subtracting a satellite PSF that is scaled by a spectrum extracted by PSF photometry. 

If the `satellite' option in \starkiller\ is set to `True', it will search for satellite streaks, and if one or more is presents, fit and subtract the streaks. 
All functions used for this purpose are contained in the \texttt{sat\_killer} class. 
We use \texttt{opencv} to detect satellite streaks through the following process:

\begin{enumerate}
    \item Calculate a detection threshold: the image median plus $15\times$ the standard deviation of the image.
    \item The image is conditioned on the detection threshold and then dilated with a $9\times9$ kernel of ones using \texttt{cv2.dilate}.
    \item The dilated image is passed through the cv2 Canny edge detection algorithm \citep{Canny1986}.
    \item The cv2 Hough Line Transform \citep{hough1962method,duda1972use,ballard1981} is applied to the edge image with a vote threshold of 100, minimum line length of 100~spaxels, and maximum line gap of 50~spaxels. 
    While these parameters are sufficient for MUSE, fine tuning may be required for other IFUs.
    \item Lines are grouped by finding the average distances between all points in each line to every other line. 
    If the average line distance is less than a maximum separation distance, they are grouped.
    \item The final consolidated line is then used to calculate the streak center, length, and angle, which are key parameters used in PSF fitting.
\end{enumerate}

If one or more satellite streaks are identified, \starkiller\ then fits a model PSF to each satellite streak. 
Currently, \starkiller\ assumes that each satellite streak can be modeled as an extremely streaked source, and so uses the best-fitting model PSF parameters determined when earlier fitting the stars in the field. 
The satellite PSF is then constructed using the star PSF parameters and the line properties.
Since satellite sky locations are frequently highly uncertain due to maneuvers and drag effects, satellite shape models and reflectance functions are infrequently made public, and satellite materials are almost always kept proprietary, we lack a comprehensive database of satellite  spectra, with efforts ongoing \citep[e.g.][]{Battle2024}. 
We therefore cannot forward-model the satellite spectrum as we do with the stars. 
Instead, we simply fit a single flux value for each wavelength, through basic PSF photometry with the satellite PSF. 
While simple, this method limits the influence of astrophysical sources on the extracted satellite spectrum, as in most fields they will only occupy a small fraction of the satellite's trail length, and thus not be favored in the fit. 

\begin{figure*}
\centering
\includegraphics[width=\textwidth]{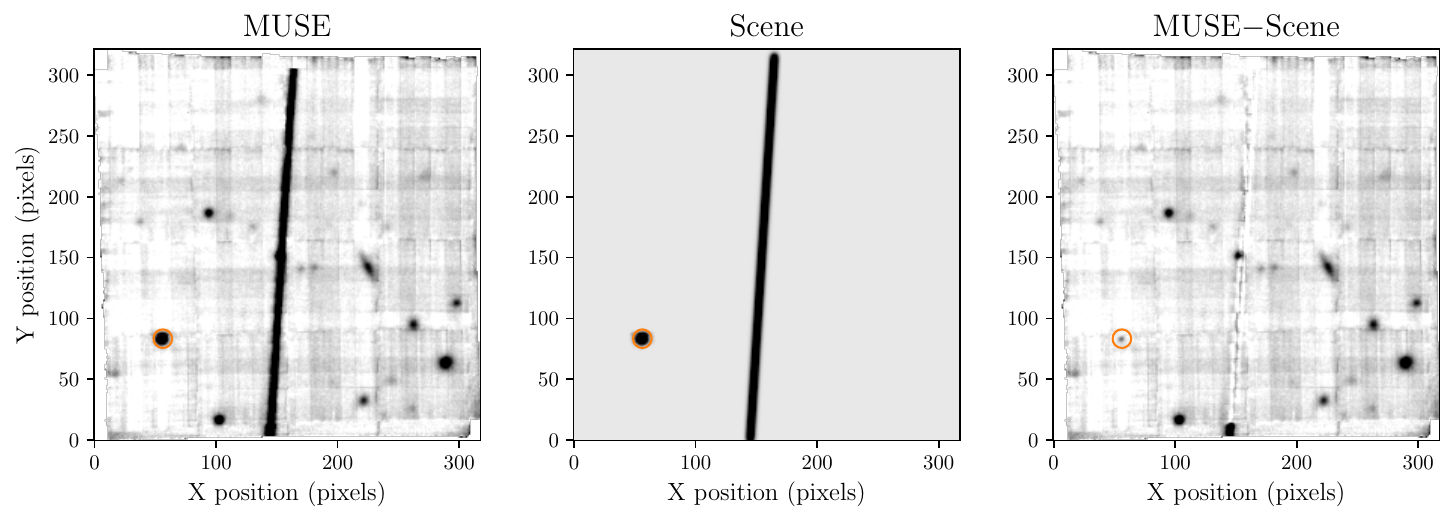}
\caption{\starkiller\ applied to the datacube where a satellite strikes Blazar WISEA J014132.24-542751.0 (J0141-5427) on 2022-Jul-23. 
(Left) In a worst-case scenario satellite strike, the unknown satellite crosses directly over the central science target. 
With the \texttt{satkiller} extension to \starkiller\ we are able to effectively model (Middle) and remove the satellite signal in all wavelengths (Right), such that the residuals are on the same scale as the detector imperfections. With \starkiller\ we have potentially salvaged a 2960~s exposure from a satellite strike. The orange circle is a field star identified by the \textit{Gaia} DR3 catalog. 
\label{fig:satellite_subtraction}}
\end{figure*}

An example of \starkiller\ applied to a satellite streak is shown in Fig.~\ref{fig:satellite_subtraction} for a MUSE observation centered on Blazar WISEA J014132.24-542751.0 (J0141-5427). 
In this worst-case scenario satellite strike, the unidentified satellite crosses directly over the science target. 
With \starkiller, we are able to effectively model this satellite streak and subtract it from the MUSE datacube, potentially salvaging a 2960~s exposure. 
We discuss a comparison of the contaminated and clean spectra of J0141-5427 in \S~\ref{sec:sat_impacts}.

\subsection{Simulated datacube construction and subtraction}
With the PSF defined, and all catalog sources matched with model stellar spectra, which is independently flux calibrated, \starkiller\ can generate a `scene' of the input datacube. 
The simulated cube is generated at 10 times the spatial resolution of the input cube, to allow for fine positioning of sources. 
We also extend the $x$ and $y$ spatial dimensions of the scene by the trail length, to include sources that are partially contained in the observed datacube.

For every astrophysical source, we create a `seed' image, by convolving the super-sampled PSF model with the position of the source in the image. 
These seed images are then multiplied with their respective stellar spectral models to create a simulated target. 
If satellites are detected in the image, they are added through the same process; however, the seed is multiplied by the PSF spectrum from the IFU. 
All sources are then combined to create the final \starkiller\ scene, and finally, the scene is downsampled to match the input cube dimensions. 
Examples of the final scenes are shown in the middle panels of Figs.~\ref{fig:subtraction},~\ref{fig:nebula_subtraction}~\&~\ref{fig:dart}.

Once the \starkiller\ scene is generated, it is subtracted from the observed datacube. 
The resultant cube is saved as a FITS file, alongside diagnostic figures for the matched spectra.

\section{Discussion}
\label{sec:discussion}

With \starkiller, we have developed a new method of analyzing crowded IFU data to primarily aid in the analysis of non-sidereally tracked extended sources. 
This technique allows us to improve crowded data such that exposures that would have otherwise been dropped can now be included in the data analysis. 
While PSF extraction pipelines such as PampelMUSE can provide precise PSF subtractions and photometry for sidereally tracked data, \starkiller\ is applicable to both sidereal and non-sidereal tracked data. 
Since both pipelines can be applied to sidereal data, we compare them in Appendix~\ref{sec:comparison}.
Furthermore, \starkiller\ opens up the possibility to perform difference imaging of IFU data cubes with single exposures, while retrieving estimated stellar parameters.

\subsection{\starkiller\ capabilities and potential uses}

\begin{figure*}[h]
\centering
\includegraphics[width=\textwidth]{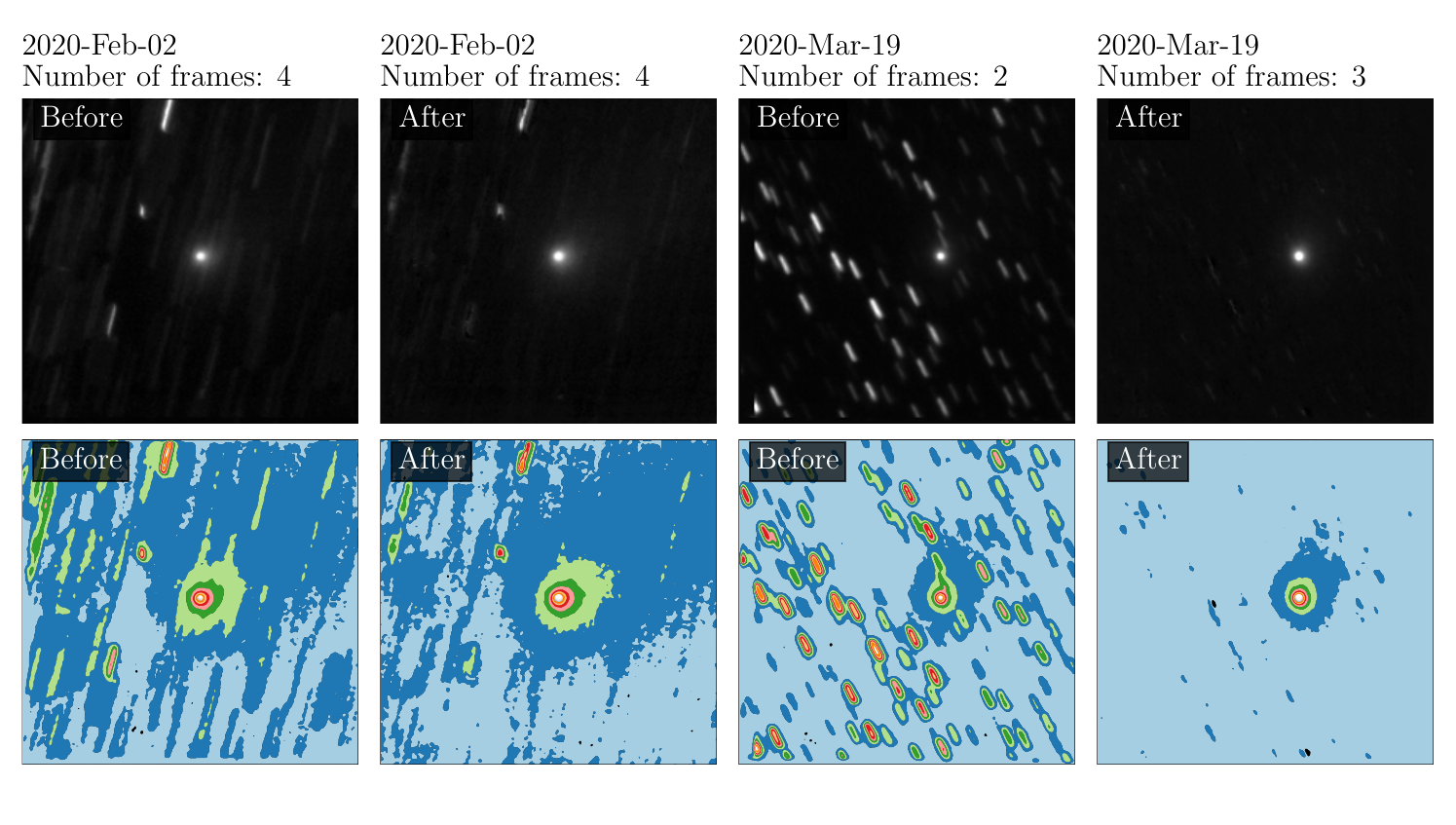}
\caption{ Images of the dust emission (7080\AA{}--7120\AA{}) from 2I/Borisov \citep{Deam:2024} before and after the application of \starkiller\ to MUSE datacubes. 
2I was within 10 degrees of the Galactic plane on 2020-Feb-02 and 2020-Mar-19, and the FOV for each exposure contained between 80 and 136 stars with m$_{GAIA-G} < 21$. 
More significant improvement occurred for the 2020-Mar-19 data: the number of usable exposures (without a star directly behind 2I) increased from 2 to 3 due to \starkiller, allowing for improved median co-adding and the removal of more stars. The greater number of bright stars in the March observations allowed for more accurate modeling of the streaked PSFs and their subsequent subtractions.
The upper images are displayed with a square root stretch over a 99.95 percentile interval, while the lower images show contours from a Gaussian convolved image of 1 standard deviation.
\label{fig:2I_final_improvement}}
\end{figure*}

\begin{figure*}
\centering
\includegraphics[width=\textwidth]{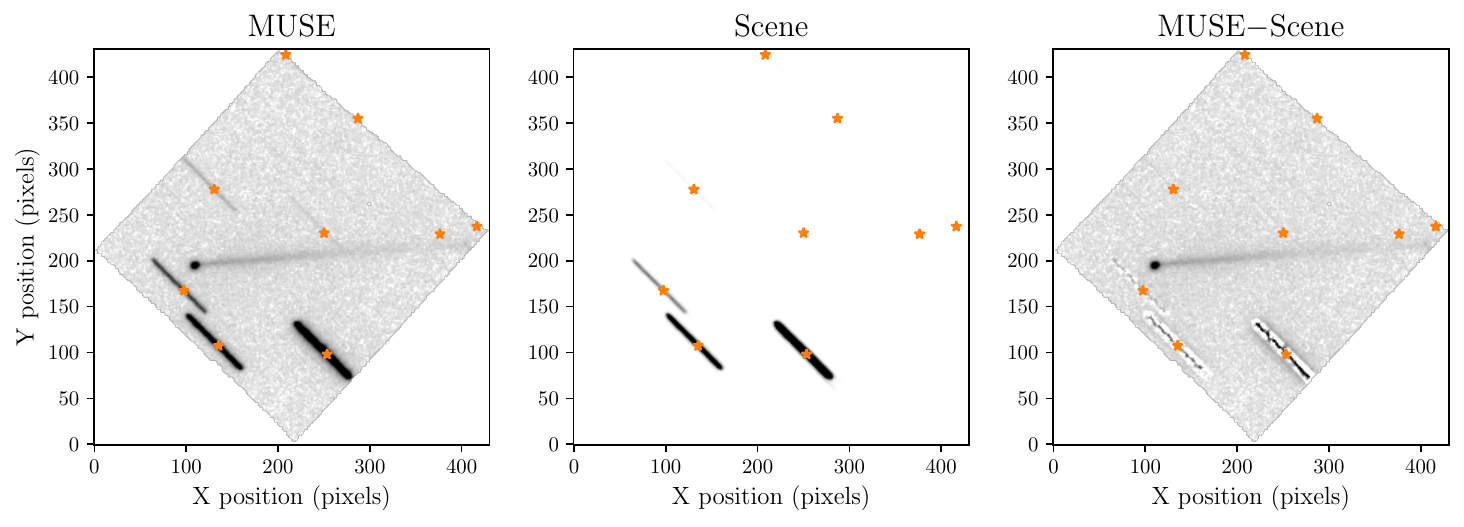}
\caption{\starkiller\ applied to a MUSE datacube of Didymos, observed on 2022-10-25, following the DART impact of 2022-09-27 \citep{Opitom:2023}. 
Faint stellar sources crossing the tail are well-modeled and subtracted. 
Brighter sources have poor subtractions in the wings of the PSF; this limitation is discussed in Sec.~\ref{sec:limitations}.
\label{fig:dart}}
\end{figure*}

The development case for \starkiller\ was the analysis of MUSE datacubes for 2I/Borisov, which are presented in \citet{Deam:2024}. 
Many of these observations were within 10 degrees of the galactic plane, where star crowding heavily limited the data quality. 
Of the 51 exposures that were taken during the 2I/Borisov observing campaign, \starkiller\ needed to be applied to 27. 
The reduction improved the data quality for dust/gas maps of 2I's coma in 23, and in the remaining 4, \starkiller\ improved the data quality so significantly that the exposures no longer had to be rejected from analysis \citep{Deam:2024}, recovering data for an unusual transient target.
Similarly, applying \starkiller\ provides improved detail in the debris trail from Dimorphos of \citet{Opitom:2023,Murphy:2023}.
This is demonstrated in Fig.~\ref{fig:2I_final_improvement} and Fig~\ref{fig:dart} respectively.
\citet{Murphy:2023} had to discard a significant portion of their debris trail datapoints, due to contamination from the stars crossing the tail.
In theory, \starkiller\ means the Solar System community can now use any IFU to look near the Galaxy.

For serendipitously observed Solar System objects, \starkiller\ provides the capacity to return time-resolved spectra on streaks.
For instance, the foreground of asteroids present in many long-duration sidereally tracked exposures can be extracted (using the \texttt{sat\_killer} mode).

\begin{figure*}
    \centering
    \includegraphics[width=\textwidth]{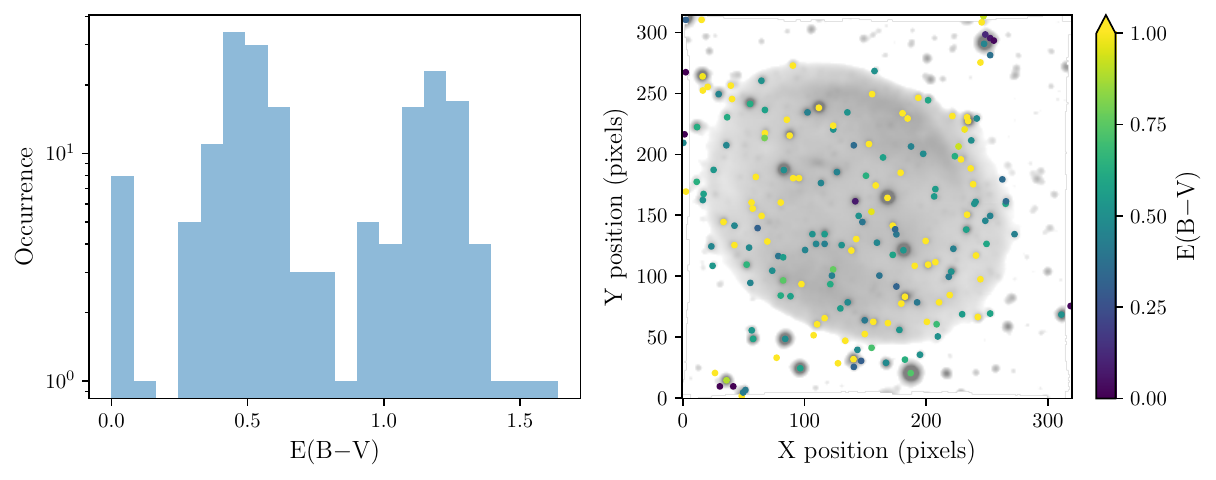}
    \caption{Best fitting E(B$-$V) values for the \textit{Gaia} DR3 stars in NGC~6563. (left) Histogram of E(B$-$V) occurrence rates, (right) spatial distribution superimposed on the IFU median image. The bimodal distribution may indicate a population of low-extinction foreground stars, and a population of high extinction stars which are largely co-located with the nebula. While the validity of extinctions that \starkiller\ applies to models is yet to be tested, it provides a way to independently and rapidly calculate extinction values for stars in a cluster.}
    \label{fig:extinction}
\end{figure*}

Since \starkiller\ identifies the best fitting extinction E(B$-$V) for each source, it may be possible to use it to calculate, or to independently check, the extinction present in star clusters. 
We present in Fig.~\ref{fig:extinction} the distribution of extinction values obtained by \starkiller\ for NGC~6563 (Fig.~\ref{fig:nebula_subtraction}) as an example for this use case. 
The distribution of extinctions present in this field is highly bimodal, displaying low and high extinction populations with median E(B$-$V) of $\sim 0.5\;$mag, $\sim 1.2\;$mag, respectively. As the high extinction population is largely co-located with NGC~6563 this may suggest that they are being obscured by the nebula. While these are Gaia DR3 sources, most do not have distances to compare against. Furthermore, we note that the E(B$-$V) value of the low extinction population is approximately double the S\&F dust map for the region, which has $\rm E(B-V)=0.2271\pm0.0024$\footnote{Accessed through: \url{https://irsa.ipac.caltech.edu/applications/DUST/}} \citep{Schlafly2011}. The large discrepancy in values for this case requires further investigation to ascertain the reliability of extinction values generated by \starkiller.
As with the flux calibration process, \starkiller\ is only optimizing the correlations between individual models and the observed spectra. 
Therefore, if the input stellar atmosphere library is insufficient, \starkiller\ may be using extinction as a tool to reshape poorly matched spectra to improve the correlation.

In \starkiller's stellar fitting, the point-source fluxes are quantified according to the expected \textit{Gaia} catalog magnitudes.
Where fully stellar PSFs remain as residuals, this can be used to identify variable sources, a common practice in conventional difference imaging.

\subsection{Satellite impacts on spectroscopic data}
\label{sec:sat_impacts}

The growing number of satellites in low-Earth orbit present fundamental challenges for all ground-based observing. 
As the number of satellites approach the projected numbers in the variety of orbital shells, their density will be sufficient such that even instruments with small FoVs will have frequent satellite streaks. 
With \starkiller, we provide a tool to isolate and remove satellite streaks from IFU datacubes. 
Furthermore, we can directly analyze a satellite spectrum, and even test for temporal variability. 
This initial approach can be expanded to become a robust method for addressing satellite contamination, and the generation of a library of satellite spectra. 

\begin{figure*}
    \centering
    \includegraphics[width=\textwidth]{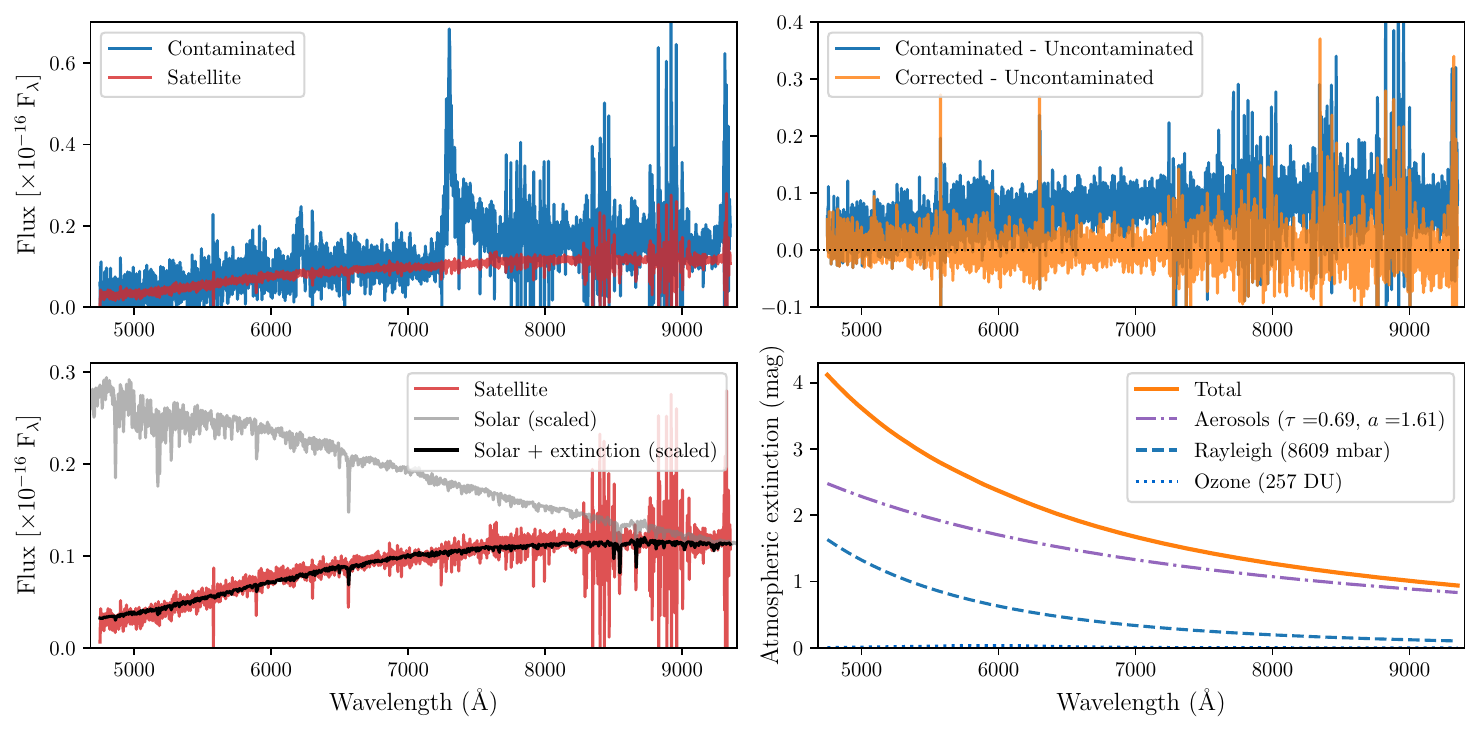}
    \caption{Extracted spectra of the satellite strike on Blazar WISEA J014132.24-542751.0 (J0141-5427) on 2022-Jul-23 (see Fig.~\ref{fig:satellite_subtraction}).
    {\bf Top left:} Blazar with satellite contamination (blue), with the satellite spectrum extracted from its streak (red) by \starkiller's \texttt{sat\_killer} functionality.
    {\bf Top right:} The residuals from subtracting another MUSE spectrum (`uncontaminated') of the blazar that was acquired immediately after the satellite passed. (blue) equivalent to the satellite spectrum (i.e. red line in top left panel); (orange) the formerly contaminated blazar data after \starkiller\ was applied. 
   \textbf{ Lower left:} The satellite spectrum (red) compared with a solar spectrum (gray) and the best-fitting atmospheric extinction for a solar spectrum (black). Several prominent solar lines are visible in the satellite's spectrum, such as H$_\alpha$ and the Na doublet, but others are not Solar or telluric.
    \textbf{Lower right}: Best-fitting atmospheric extinction that maximizes the correlation of an extincted solar spectrum to the satellite spectrum (orange), and the three dominant extinction components from aerosols (purple dot-dash), Rayleigh scattering (blue dashed), and ozone (blue dotted). }
    \label{fig:sat_spec}
\end{figure*}

Satellite streaks directly crossing science targets, or `strikes' in IFU data provides us with a unique opportunity to examine how satellite strikes impact optical observations, which has not yet been shown for MUSE. 
Fig.~\ref{fig:sat_spec} demonstrates that the satellite observed in the datacube of J0141-5427 has a spectrum that resembles a smooth continuum, with some prominent absorption lines visible in Fig.~\ref{fig:sat_spec} (top left). 
Both this unidentified satellite and the blazar are SNR $\sim 100$, generating a large effect on the blazar spectrum.
This appears more severe than the expectations for spectroscopic impact of \citet{Hainaut:2024}.
However, in comparison with a spectrum of J0141-5427 that was observed subsequently and is without a satellite passage, we find that \starkiller's satellite subtraction process successfully removes the broad satellite continuum, as seen with the orange line in Fig.~\ref{fig:sat_spec} (top right). 
In this case, we believe that \starkiller\ has effectively `cleaned' the datacube of the satellite contaminant, allowing it to now be used for science.

With this well-isolated satellite spectrum, we can analyse its properties and the atmospheric effects. 
As the satellite is reflecting sunlight, we might expect the satellite spectrum to be similar to the Sun; this is a frequent assumption in studies of potential satellite impact on astronomical spectra \citep[e.g.][]{Bialek:2023,Hainaut:2024}.
As shown in Fig.~\ref{fig:sat_spec} (lower left), we find that the satellite spectrum is indeed well represented by a Solar spectrum that has been highly processed by atmospheric extinction, with the parameters fit by \texttt{pyExtinction} \citep{pyExtinction}. 

While the MUSE pipeline applies a correction for standard atmospheric extinction \citet[][Sec.~4.9]{Weilbacher2020}, it is possible that the satellite has experienced higher levels of atmospheric extinction. For satellites, the relative configuration of the Sun, satellite, Earth, and observer become important. In many configurations it is possible for light from the Sun to pass through the Earth's atmosphere before reaching the satellite. This additional passage of light through the atmosphere would create the enhanced atmospheric extinction that we observe in the satellite spectrum.

\subsection{Independent flux calibration verification}

Through the forward-modeling approach of \starkiller, we have developed a way to independently verify the flux calibration of an IFU datacube. 
As described in Sec.~\ref{sec:spec_match}, we only use the observed spectra for shape comparison by correlating the observed spectra with model spectra. 
The best-fitting model spectra are then flux scaled according to catalog magnitudes (\textit{Gaia} DR3 in the default case); we then compare the flux of these scaled models to the observed spectra to generate a flux correction, as described in Sec.~\ref{sec:flux_correction}. 
While some fine features in the flux correction may result from poor model matching, overall trends and flux offsets are likely to be indicative of a flux calibration error between the datacube and the catalog. 

\subsection{Current \starkiller\ limitations}
\label{sec:limitations}

While \starkiller\ is robust to a wide range of challenging observational conditions, there are several limitations. 
Primary among these is the source density: if there are too few stars in the FOV or in the source catalog, then \starkiller\ is unable to create corrections based on the population of sources. 
If there are few sources i.e. $<3$ sources contained in the IFU, then the data PSF and flux correction will be biased by the available targets. 
Flux correction will only be applied if there are at least 3 calibration sources present in the IFU, or the \texttt{force\_flux\_correction} option is set. 
If the catalog is incomplete, then un-cataloged sources will not be subtracted: in the example of NGC~6563 (Fig.~\ref{fig:nebula_subtraction}), the majority of sources that can be seen are not identified in the \textit{Gaia} DR3 catalog. 
Additionally, the selected calibration sources may then face higher-than-expected crowding. 
Unexpected crowding of calibration sources can lead to poor PSF models, and therefore poor subtractions. 

Conversely, while \starkiller\ is fairly robust to crowding, overlapping sources are not well modeled. 
The current method for spectral extraction and source position fitting considers each source independently. 
For crowded sources, this approach leads to spectral contamination and poor positional fitting for crowded sources, particularly faint sources.
An alternative approach such as iterative PSF fitting and subtraction of targets, or simultaneous fitting of grouped sources, could be more successful for overlapping sources. 
In cases of sidereal tracking of crowded sources, PampelMUSE will provide higher quality spectra and PSF subtractions.

The \starkiller\ PSF construction method currently does not create wavelength-dependent PSFs. 
While this limitation is largely inconsequential for the low signal-to-noise stars in non-sidereal observations, it may limit the reliability of \starkiller\ in high signal-to-noise observations of bright stars in sidereal observations, a situation where PampelMUSE is better suited.

Another limitation of the current PSF implementation is how the PSF wings are treated, particularly for the data PSF. 
To avoid including detector artifacts into the data PSF, we set all spaxels with less than $10^{-4}$\% contribution to the total PSF to 0. 
This sets a nominal PSF radius in MUSE data to be $\sim8\;$spaxels. 
For faint sources, this PSF truncation has little effect, as the flux at the PSF wings is low; however, for bright sources, this leads to overly-subtracted centers, surrounded by un-subtracted wings. 
A possible solution to this would be to augment the data PSF with a Moffat or Gaussian model component fit to the PSF wings. 

The quality of the \starkiller\ subtraction for a target depends heavily on how well the spectral models fit. 
While we have equipped \starkiller\ with a broad sample of stellar spectra, additional models may be required, or be more representative of specific stars. 

Even with these limitations, \starkiller\ is effective at removing more than 90\% of the stellar flux in most cases. 
For non-sidereal targets, the combination of \starkiller\ and averaging together multiple cubes observed at different sky locations overcomes much of the limitations. 
Some refinement is still required to optimize \starkiller\ for sidereal-tracked observations with high signal-to-noise stars.

\subsection{Extending \starkiller\ to other IFUs}
\label{sec:other_ifus}

With a catalog in hand, \starkiller\ can be extended to other IFUs; currently operational ones have smaller FoVs than MUSE.
For instance, it would be immediately suitable for WiFes (38\arcsec$\times$25\arcsec), GMTIFS (20.4\arcsec$\times$20.4\arcsec) or Keck KCWI (20\arcsec$\times$33\arcsec) data that has the same data format as MUSE.
Planned ELT IFUs (e.g. HARMONI, WFOS) have similar-scale FOVs. 
Even for the smallest IFUs, the capabilities of \starkiller\ will be suited for treating satellite-affected data.
The future VLT instrument BlueMUSE\footnote{\url{https://bluemuse.univ-lyon1.fr/} } is planned to have a FOV larger than 1 arcmin$^2$, which would be highly suited to \starkiller.

For JWST's NIRSpec or MIRI, we note that the default stellar models in \starkiller\ are cut to the optical range; similarly, using an IR catalog would be necessary.

In all cases, for fast-moving non-sidereal objects, in the present version of \starkiller, the exposure lengths of observations would need to be capped, so as to retain the star trail length within the IFU FOV (the containment requirement discussed in \S~\ref{sec:psf}).

\section{Conclusion}

In this paper we have presented the \starkiller\ package, which creates synthetic difference images of single IFU data cubes using a forward-modeling approach. 
By utilizing independent photometric catalogs, and a suite of stellar atmosphere models, \starkiller\ simultaneously provides stellar spectral classification, relative velocity, and line-of-sight extinction for all sources in a catalog, alongside a source-subtracted datacube.

We developed \starkiller\ to be compatible with both sidereal and non-sidereally acquired observations. 
For streaked sources, we do not require input tracking, as \starkiller\ is generalized to work with even extreme cases of elongated sources and crowded fields. 
In the most extreme case, it can model and subtract satellite streaks.
We developed this method to clean highly elongated stars from VLT/MUSE observations of 2I/Borisov. 
As we were working to preserve the spectral features of a diffuse foreground object, we developed \starkiller\ to rely heavily on catalogs of model stellar spectra and magnitudes.

IFUs provide exceptional capabilities for astronomical enquiry.
We look forward to seeing what uses this package may find in the community.

\newpage
\begin{acknowledgments}


R.R.H. is supported by the Royal Society of New Zealand, Te Ap\={a}rangi through a Marsden Fund Fast Start Grant and by the Rutherford Foundation Postdoctoral Fellowship.
M.T.B. appreciates support by the Rutherford Discovery Fellowships from New Zealand Government funding, administered by the Royal Society Te Ap\={a}rangi.

Package development based on observations collected at the European Southern Observatory under ESO programmes 
103.2033.001--003 and 105.2086.002 (2I/Borisov, PIs: M.T.B and Cyrielle Opitom), 
110.23XL and 109.2361 (DART, with thanks to PI Cyrielle Opitom), 
60.A-9100 (PI: MUSE Team), 
and 109.238W (PI: Fuyan Bian). 
We thank the ESO staff, particularly Danuta Dobrzycka and Lodovico Coccato of the MUSE SDP team, Marco Berton, Henri Boffin, Bin Yang, Diego Parraguez, Edmund Christian Herenz, Fuyan Bian, and Israel Blanchard, for their help in the acquisition or retrieval of these observations.

We thank the community of the IAU Centre for Protection of the Dark and Quiet Sky From Satellite Constellation Interference for their help during this work.

This work has made use of data from the European Space Agency (ESA) mission
{\it Gaia} (\url{https://www.cosmos.esa.int/gaia}), processed by the {\it Gaia}
Data Processing and Analysis Consortium (DPAC,
\url{https://www.cosmos.esa.int/web/gaia/dpac/consortium}). Funding for the DPAC
has been provided by national institutions, in particular the institutions
participating in the {\it Gaia} Multilateral Agreement.

\end{acknowledgments}

\vspace{5mm}
\facilities{VLT(MUSE)}

\software{\texttt{Starkiller} \citep{starkiller_code},
astropy \citep{Astropy2013,Astropy2018,Astropy2022}, 
astroquery \citep{astroquery},
scipy \citep{2020SciPy-NMeth},
photutils \citep{photutils},
pandas \citep{mckinney-proc-scipy-2010,reback2020pandas},
numpy \citep{numpy},
matplotlib \cite{Hunter2007},
OpenCV \citep{opencv_library},
MUSE Python Data Analysis Framework \cite[MPDAF;][]{MPDAFsoftware},
TRIPPy \citep{Fraser:2016},
pyExtinction \citep{pyExtinction},
PampelMUSE \citep{Kamann:2013}
          }

\newpage
\appendix
\section{PSF modeling of point sources} \label{sec:ps_comparison}
\begin{figure*}
\centering
\includegraphics[width=\textwidth]{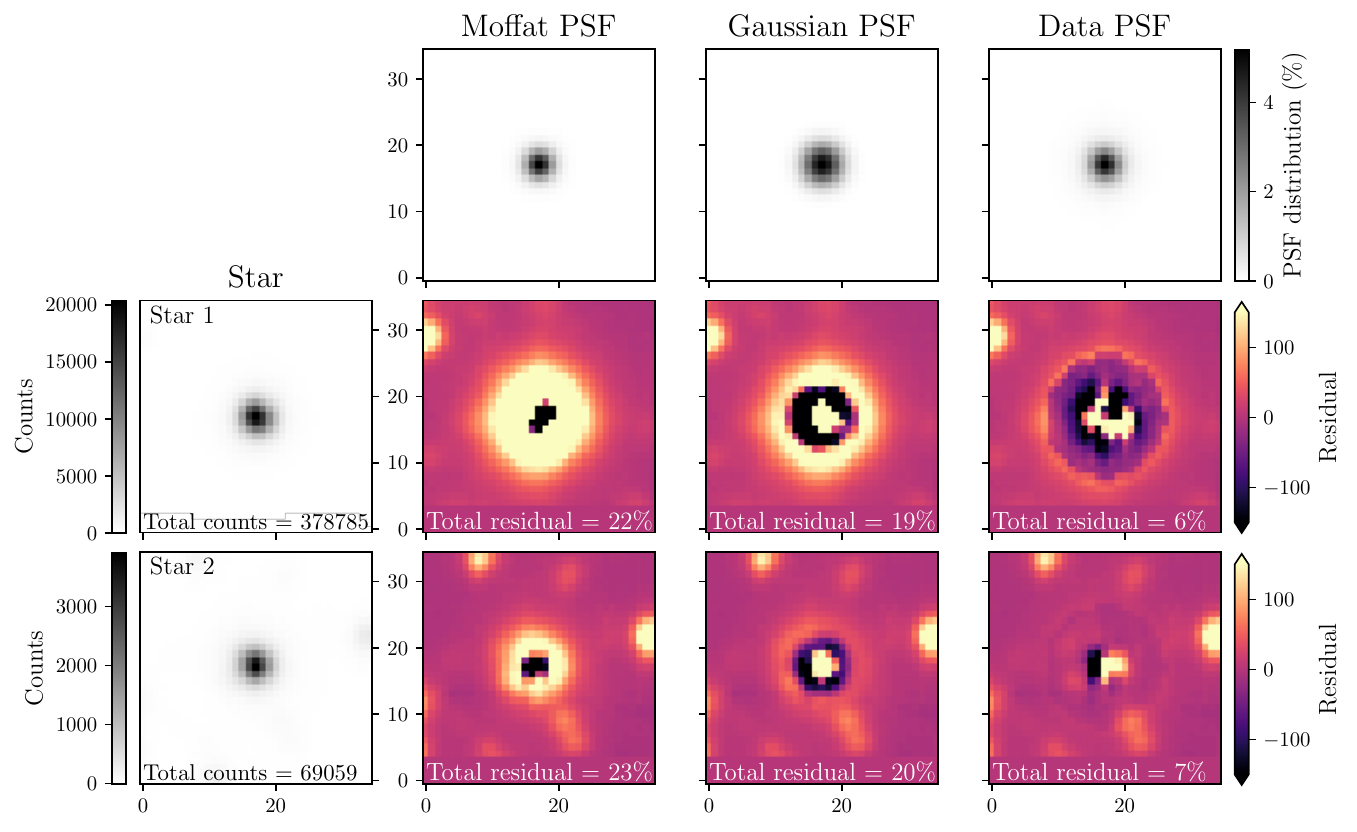}
\caption{Recreation of Fig.~\ref{fig:psf_fit} for sidereally tracked point sources extracted from the sky observation ADP.2023-09-27T15\_07\_36.629. 
As with the trailed sources, the data PSF outperforms both the Moffat and Gaussian PSF profiles.
\label{fig:psf_fit_ps}}
\end{figure*}

The functions \starkiller\ uses to model trailed PSFs can be readily applied to point sources. If the optional variable \texttt{trail} is set to false, the trail length will be set to 1 spaxel, and \starkiller\ will fit regular point sources. As seen in Fig.~\ref{fig:psf_fit_ps}, the data PSF outperforms both the Gaussian and Moffat PSF with residuals $\sim15\%$ lower for the data PSF.

\section{Comparison of sidereal source-subtraction with PampelMUSE} \label{sec:comparison}
The PempelMUSE pipeline \citep{Kamann:2013} was developed to obtain high precision extraction of point sources from MUSE data. 
Some aspects of PampelMUSE are similar to \starkiller, such as requiring a source catalog to identify sources; both pipelines can produce source-subtracted cubes. 
Unlike \starkiller, PempelMUSE can only be applied to sidereally tracked data, and primarily uses a Moffat profile to model the PSF. 
PempelMUSE creates its source-subtracted data cube through subtracting the fit PSF profile for each source from every frame. 
Therefore, while the two pipelines were created for different use cases, we can draw comparisons between the two when looking at their outcomes on sidereally tracked data. 

\begin{figure*}
\centering
\includegraphics[width=\textwidth]{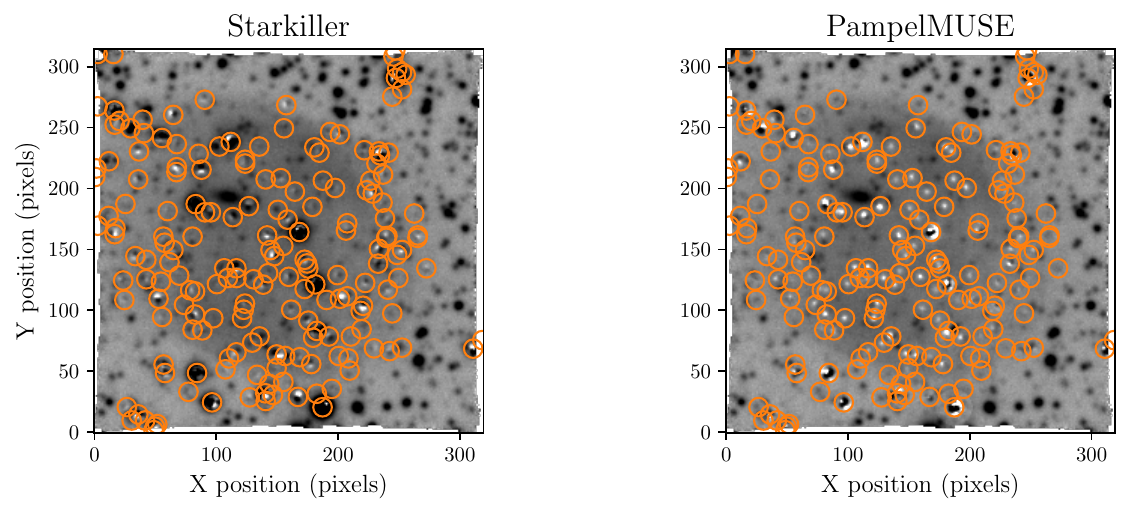}
\caption{Comparison of source-subtracted MUSE datacubes for NGC~6563 processed by \starkiller\ (left) and PempelMUSE (right). The two pipelines take different approaches: \starkiller\ uses spectral models scaled by external flux catalogs, and PempelMUSE directly subtracts sources through PSF fitting. In this example, PempelMUSE appears to systematically over-subtract sources, particularly those that overlap with the nebula.
\label{fig:sk_pamp_comp}}
\end{figure*}

In Fig.~\ref{fig:sk_pamp_comp} we compare the subtracted cubes produced by \starkiller\ and PempelMUSE for NGC~6563 (the same nebula shown in \S~\ref{sec:sidereal} and Fig.~\ref{fig:nebula_subtraction}). 
While both model have subtraction artifacts, PempelMUSE systematically over-subtracts sources within the extent of the nebula. 
This over-subtraction will result in output spectra from PempelMUSE that have larger fluxes than were observed.
In contrast, \starkiller\ appears to under-subtract sources, which will be due to the flux scaling to match the catalog photometry and any intrinsic variability of sources. 
PampelMUSE outperforms \starkiller\ for crowded sources. 
This test also highlights the poor subtractions from \starkiller\ and PampelMUSE for bright sources, suggesting that the MUSE PSF profile differs between bright and faint sources.

\bibliography{refs}{}
\bibliographystyle{aasjournal}

\end{document}